\documentclass[twocolumn,amsmath,aps,nofootinbib,superscriptaddress]{revtex4}

\usepackage{amssymb}
\usepackage{amsmath}
\usepackage{epsfig}
\usepackage{subfigure}
\usepackage{mathrsfs}
\usepackage{longtable}
\usepackage{epstopdf}

\newcommand{\be}{\begin{equation}}
\newcommand{\ee}{\end{equation}}
\newcommand{\bea}{\begin{align}}
\newcommand{\eea}{\end{align}}

\newcommand{\unit}[1]{\ensuremath{\, \mathrm{#1}}}

\begin{document}

\title{Ultra-light Axions: Degeneracies with Massive Neutrinos and Forecasts for Future Cosmological Observations}
\author{David J. E. Marsh}
\email{d.marsh1@physics.ox.ac.uk}
\affiliation{Rudolf Peierls Centre for Theoretical Physics, University of Oxford, 1 Keble Road, 
Oxford, OX1 3NP, UK}
\author{Edward Macaulay}
\email{e.macaulay1@physics.ox.ac.uk}
\affiliation{Astrophysics, University of Oxford, DWB, Keble Road, Oxford, OX1 3RH, UK}
\author{Maxime Trebitsch}
\email{maxime.trebitsch@ens-lyon.org}
\affiliation{D\'{e}partement de Physique, \'{E}cole Normale Sup\'{e}rieure de Lyon, 46 all\'{e}e d'Italie, 69364 LYON cedex 07}
\author{Pedro G. Ferreira}
\email{p.ferreira1@physics.ox.ac.uk}
\affiliation{Astrophysics, University of Oxford, DWB, Keble Road, Oxford, OX1 3RH, UK}


\begin{abstract}
A generic prediction of string theory is the existence of many axion fields. It has recently been argued that many of these fields should be light and, like the well known QCD axion, lead to observable cosmological consequences. In this paper we study in detail the effect of the so-called string axiverse on large scale structure, focusing on the morphology and evolution of density perturbations, anisotropies in the cosmic microwave background and weak gravitational lensing of distant galaxies. We quantify specific effects that will arise from the presence of the axionic fields and highlight possible degeneracies that may arise in the presence of massive neutrinos. We take particular care understanding the different physical effects and scales that come into play. We then forecast how the string axiverse may be constrained and show that with a combination of different observations, it should be possible to detect a fraction of ultralight axions to dark matter of a few percent. 
\end{abstract}

\maketitle
\section{Introduction: The Axiverse and Cosmology}
\label{intro}

There is a widespread consensus that the Universe can be accurately described by
General Relativity and the statistical physics of particles and fields. The quantitative
model to arise from such a description can be used to accurately predict the
gross features of the Universe (such as, for example, its expansion rate and the spectrum of the cosmic microwave background)  and some detailed characteristics (such as, for example, the abundance of light elements, the number of relativistic species, the density of baryons and the curvature of space). In fact, 
over the past few years, cosmological observations have led to an ever increasing number
of precision constraints on a variety of cosmological parameters that can describe the very early Universe,
the evolution of the Universe over many orders of magnitude in scale and the current state of the Universe. Again and again, the favoured cosmological model has passed observational tests with flying colours. It has been dubbed the {\it concordance model} \cite{book:dodelson}.

The concordance model is incredibly robust yet, at the same time, it predicts a whole new
sector of fundamental physics which has yet to be understood. In the concordance model,
$96\%$ of the Universe is dark, that is, doesn't interact or only interacts very weakly with electromagnetic radiation. This dark sector divides up into two components: roughly $20\%$ of it is in the form of dark matter, a gravitationally interacting type of pressureless matter which can clump, and $76\%$ in the form
of dark energy, which is gravitationally repulsive and can drive the accelerated expansion of the Universe
at late times. The overriding goal of modern cosmology is now to understand the dark sector using the tools
of what has been called {\it precision cosmology} \cite{7yearWMAP}.

The focus in the quest for dark matter and dark energy is to model and measure the large scale structure of the Universe.
By this we mean the distribution of galaxies as a function of redshift, the anisotropies of the cosmic microwave background and the weak gravitational lensing of distant galaxies. The progress in measuring
these different observables has been tremendous and will continue, with new experiments either being planned or coming online over this coming decade. 

Given the promise of observational cosmology, the hope is to not only understand the nature of dark energy and dark matter but also explore other aspects of the concordance model and, in particular, measure in more detail other parts of the matter and energy content of the Universe. One obvious
component of interest is the relativistic dark matter sector in the form of, for example, massive neutrinos. Another possibility is the existence of ultra-light scalar fields that arise in some versions of string
cosmology in what is known as the {\it string axiverse}~\cite{axiverse2009}. Both of these components are intimately tied to
the dark sector and will be the focus of this paper.

\subsection*{Additional Ingredients in the Hot or Relativistic Dark Matter Sector}
\label{introduction_ingredients}

The natures of dark matter (DM) and dark energy (DE) are completely unknown, and theorists argue intensely even about their very existence, yet their relative necessity in understanding precision cosmological data has made their gross properties in terms of $\Omega_d$, $\Omega_\Lambda$, and to a lesser extent the dark energy equation of state parameters $w_0$ and $w_a$, become an integral part of the concordance model. In addition to these and the other established ``vanilla'' cosmological parameters, there is mounting evidence for additional cosmological ingredients, possibly coming from multiple sources.

Observations of the cosmic microwave background (CMB) and large scale structure (LSS) in recent years have consistently found the need for excess relativistic energy density \cite{hamann2007,hamann2010b,gonzalez-garcia2010,7yearWMAP,dunkley2010} \footnote{However, the authors of \cite{gonzales-morales2011} suggest that this may be due to priors}. This excess radiation is parameterised in terms of the effective number of relativistic neutrino species, $N_{\mathrm{eff,rel}}$ \cite{shvartsman1969,steigman1977,linde1979}, as:
\begin{equation}
\rho_R = \left[ 1+\frac{7}{8}\left( \frac{4}{11} \right)^{4/3} N_{\mathrm{eff,rel}}\right] \rho_\gamma
\end{equation}
where $\rho_\gamma$ is the energy density in photons fixed by the CMB temeperature. Even within the standard model of particle physics, with no neutrino masses, $N_{\mathrm{eff,rel}}$ can be non-integer and greater than three if neutrino decoupling is non-instantaneous and the thermal neutrinos are partially reheated by electron positron annihilation \cite{dolgov2002}. It is often stated that three massless standard model neutrinos are best described by $N_{\mathrm{eff,rel}}\approx 3.04$. Any increase from this is thought of as ``extra''.

The radiation density at big bang nucleosynthesis (BBN) is constrained by the light element abundances, but bounds vary depending on the treatment of astrophysical uncertainties, new physics scenarios and improved measurements. For example, \cite{cuoco2004} has $N_{\mathrm{eff,rel}}=2.5^{+1.1}_{-0.9}$, while \cite{hamann2010a} allow up to $\Delta N_{\mathrm{eff,rel}}=N_{\mathrm{eff,rel}}-3.04=1.39$ at 95\% credible interval. However, the BBN limits need not apply at the CMB or LSS scales, since late decaying particles may increase the neutrino abundance after BBN, but before or after CMB formation. Accordingly, $N_{\mathrm{eff,rel}}$ is taken as a free parameter in most cosmological parameter estimations, with best fit values from WMAP7 $N_{\mathrm{eff,rel}}=4.34^{+0.86}_{-0.88}$ \cite{7yearWMAP}, ACT $N_{\mathrm{eff,rel}}=4.6 \pm 0.8$ \cite{dunkley2010},  SDSS-DR7 $N_{\mathrm{eff,rel}}=4.78^{+1.86}_{-1.79}$ \cite{hamann2010b}. Motivation for such extra radiation density is lacking in the standard models of particle physics and cosmology, however many theoretical extensions of these models provide clues. Such a situation is indeed expected in the ``Freeze-in'' mechanism of producing asymmetric dark matter, if the relic particles decay to neutrinos \cite{hall2009,hall2010}.

The observation of neutrino oscillations requires the introduction of neutrino masses, which constitutes a hot dark matter (HDM) component (for a review of massive neutrinos in cosmology, see \cite{lesgourgues2006}, and for a historical review of HDM see \cite{primack2001}). However the nature of neutrino masses and mixing remains unknown, and therefore so also does the exact effect of neutrino masses on cosmology. The situation is further complicated due to the degeneracy, on certain scales and for certain observables, between massive neutrinos and other cosmological parameters \cite{hu1998d}.

The situation with regards the measurement of massive neutrino parameters using terrestrial experiments is summarised in some recent fits by Giunti \cite{giunti2011}. These fits seem to favour not only massive standard model neutrinos, the absolute mass scale of which can only currently be determined by cosmology, but also the inclusion of one or two species of massive ``sterile'' neutrinos, which have a large mass splitting from the standard model neutrinos. The fits to this model and cosmology favour the standard model neutrinos being approximately massless, and the sterile neutrinos to have masses in the eV range. Sterile neutrinos are even more ambiguous cosmologically, since incomplete thermalisation allows the effective number of massive neutrinos, $N_{\mathrm{eff,mass}}$, to also take non-integer values. Cosmological fits for sterile neutrinos are given in \cite{hamann2010a}, and forecasts are made in \cite{giusarma2011}, while forecasts for the cosmological measurement of standard model neutrino mass splittings are made in \cite{bernardis2009}. It is also well known that cosmology can potentially resolve the neutrino mass hierarchy as being the ``normal'' heirarchy if the sum of neutrino masses, $\Sigma m_\nu<95\unit{meV}$, an accuracy within reach if many cosmological probes are combined \cite{joudaki2011}.

The existence of neutrino masses, while in some views of the history of particle physics strictly ``beyond the standard model'' (BSM), is at least well established enough, both experimentally and theoretically, that many would not class them as BSM at all. Their effects, as we have seen, are included in most cosmological analyses, too. Their known existence, combined with the fact that HDM cannot account for all the dark matter, is definitive evidence that the dark sector is multi-component. Another theoretically well motivated, cosmologically important ingredient that may be necessary to explain curious features within the standard model of particle physics, namely the strong $\mathcal{CP}$ problem, is the (QCD-)axion \cite{pecceiquinn1977, thooft1976a, thooft1976b, dine1981, preskill1983, steinhardt1983, turner1983, abbott1983, dine1983,ipser1983,turner1986,berezhiani1992,banks1996,fox2004,hannestad2005,visinelli2009,hannestad2010} (for reviews of axion cosmology, see \cite{sikivie2008,sikivie2010}). Depending upon their mass, axions can constitute the full range of dark matter ``temperatures'', from cold through warm to hot: a true feast for Goldilocks. A model dependent coupling to photons can make them not really dark at all, and constraints can be derived on this from dimming of supernovae \cite{mortsell2003}. Hot, thermal axions can contribute to $N_{\mathrm{eff,rel/mass}}$, but their weak couplings make this contribution fractional \cite{hannestad2005}. For our purposes, the standard axion will be considered part of $N_{\mathrm{eff}}$ or $\Omega_c$ appropriately.

Axions in the mass range $0.7\unit{eV}\lesssim m_a \lesssim 300\unit{keV}$ are excluded by cosmology for a variety of reasons \cite{hannestad2005,hannestad2010,cadamuro2011}. Sub eV mass axions contribute as HDM, and their mass is limited by constraints on $N_{\mathrm{eff}}$: in exact analogy to neutrinos, they cannot be too heavy. Heavier axions, which are too heavy to be HDM, if they couple to photons, are restricted by their decays/inverse decays via effects on  BBN, CMB distortions, and concordance between BBN and CMB determined values of the baryon to photon ratio. In addition, in this scenario, early axion decays to photons dilute the effective number of neutrino species, creating more tension with the large measured values of  $N_{\mathrm{eff}}$ quoted above. 

Goldilocks properties of axions are abundant as they make multiple changes in their dark matter temperature as one moves through their possible mass spectrum. One normally considers heavy weakly interacting massive particles (WIMPs) with GeV masses as cold dark matter (CDM), and light neutrino-like particles with eV masses as HDM, and axions can indeed populate these masses and temperatures in the same way. However, very light axions with $m\lesssim 1\mu\unit{eV}$ once again constitute CDM, and the types of limits given above from couplings to photons cease to apply \cite{sikivie2008}. As we will see through the course of this paper, another transition occurs when these light CDM-like axions become lighter still and their quantum properties cause them to behave cosmologically like HDM again. This range of behaviour has to do with competition between various physical processes which come in and out of dominance as coupling properties and relic density contributions of axions vary with mass and cosmic evolution. We do not know the fundamental model and parameters that would exactly determine axion behaviour. Names can also be deceptive and as we will see, the QCD axion is not the only axion relevant for cosmology.

In addition, from BSM particle physics there are many, many, candidates for the CDM that differ from the standard WIMP, the neutralino of the (minimally-) supersymmetric standard model, in lesser or greater ways. These differences may affect their properties during inflation, during baryogenesis, or during BBN. These early time effects are of no concern to us here. As long as the dark matter is cold, we count it into $\Omega_c$. Late time effects, such as possible decay, or couplings within the dark sector, while cosmologically relevant, are also beyond the scope of this paper.

From the point of view of non-linear structure formation and cosmological phenomenology, dark matter self-interactions \cite{spergel2000} and more novel ingredients like ``Fuzzy Cold Dark Matter" (FCDM) \cite{hu2000} have been proposed as resolutions to the problems of cuspy dark matter halos, and the large predicted but unobserved numbers of dwarf galaxies in the standard CDM model (the well known ``missing satellites'' problem). This has lead some to general consideration of Bose-Einstein-Condensate (BEC) dark matter (see, for example, \cite{silverman2002,fukuyama2008,rindler-daller2011}, and in the case of axions \cite{sikivie2009,sikivie2010b}). Indeed, the numerical simulation of \cite{woo2009} showed that the presence of such an ultra-light scalar condensate indeed reduces the number of dwarf galaxies, but in fact does very little to the cuspy density profile. However there are also many unaccounted for factors in standard galaxy formation models with CDM that may affect the formation of cusps and dwarf galaxies, such as baryon physics and supernova feedback (see, for example, \cite{pontzen2011} and references therein).

A huge amount of research in modern cosmology goes into models for the 76\% of DE that treat it other than as a cosmological constant \cite{copeland2006}, for example modified gravity \cite{clifton2011} and quintessence \cite{wetterich1988,peebles1988,ratra1988}. Many of these models alter the equation of state, $w(z)$, of DE, an effect degenerate with neutrino masses at some scales. In particular, models where there is a component of dark energy with effects at high redshift (early dark energy, EDE) are known to share many degeneracies in their effects on cosmological observables with extra relativistic energy density, massive neutrinos, and other forms of hot dark matter or any other structure-suppressing cosmological ingredients (see, for example, \cite{deputter2009,calabrese2010,joudaki2011}). The potential degeneracies can, however, be broken by the use of multiple observables \cite{joudaki2011}. We will not be studying EDE in this work, but note that any potential detection of such an exotic component can only be truly qualified if all other aspects of cosmology with potentially similar effects are well understood. Finally, massive neutrinos are a key ingredient, along with a modified inflationary period, in allowing the model of \cite{hunt2010} to fit the data and analyses of \cite{sherwin2011,hlozek2011} without the inclusion of a DE sector, so that the future success or failure of this non-standard model, too, must hinge on thorough understanding of the structure suppressing DM species.

\subsection*{The Axiverse}

Cosmologists often invoke the existence of light scalar fields in the late universe as DM and DE components, for example in theories of quintessence, coupled quintessence (e.g. \cite{amendola2000}), chameleons \cite{khoury2004,brax2004}, unified dark matter, and the Bose-Einstein condensates mentioned above. From a particle physics/string theory point of view these ingredients come up against two main problems: fifth-force constraints, and cosmologically light masses, which are not unrelated\footnote{Addressing these issues for chameleons/string moduli has been looked at in e.g. \cite{brax2007,conlon2010,hinterbichler2011}.}.

For a (coherent) scalar field to be cosmologically distinct from CDM, or to play a quintessence like role, it must be very light: $10^{-33}\unit{eV}\lesssim m_{\phi} \lesssim 10^{-18}\unit{eV}$. Gauge invariance and Lorentz invariance then allow this scalar to multiply terms in the Lagrangian of the standard model fields, leading to problematic long range ``fifth forces''. Unless a symmetry forbids them, these couplings should be universal and cannot be restricted ad hoc to the dark sector alone. If the field is to have an origin in new physics beyond the standard model then its lightness and stability also become hard to explain without introducing additional hierarchy problems. For this reason, scalars in the late universe are considered generally problematic in models of particle physics; keeping them under control is one key motivation for moduli stabilisation in string theory (see \cite{conlon2006} and references therein).

There is, however, at least one generic source of light scalars coming from high energy physics that evades all of the problems highlighted above: the so-called ``String Axiverse'' \cite{axiverse2009} (where a more detailed version of the following argument is given). Axions, as mentioned above, were first motivated to solve the ``strong $\mathcal{CP}$ problem'' of QCD, where the problematic $\mathcal{CP}$ violating parameter $\theta$ occuring in the Lagrangian as $\mathcal{L}\supset \theta \tilde{F}_{\mu\nu}F^{\mu\nu}$ is made dynamical as the Goldstone boson of a spontaneously broken global $U(1)$ symmetry, and driven to its $\mathcal{CP}$ conserving value by a potential induced non-perturbatively by QCD instantons. The quantum of excitation of $\theta$ is then the axion. The global symmetry is broken at the scale $f_a$, whilst the non-perturbative physics giving the axion its potential switches on at a scale $\mu$. This makes the axion a pseudo-Nambu-Goldstone boson (PNGB). Indeed, ultra-light axions and other PNGB's motivated in high energy physics, such as  in \cite{hill1988,hall2005}, were studied as a solution to the non-zero cosmological constant problem as early as 1995 \cite{frieman1995}, before the current vogue for quintessence in cosmology began with the supernova observations of 1998 \cite{perlmutter1999,riess1998}. 

We would like string theory to furnish us with the QCD axion and its solution to the strong $\mathcal{CP}$ problem. Axions will always arise in string theory compactifications \cite{witten1984,witten2006} as Kaluza-Klein zero modes of antisymmetric tensor (form) fields analogous to the Maxwell tensor, $F_{\mu\nu}$. These terms appear when the form fields are compactified on closed cycles in the compact space, with 3-forms being compactified on closed 3-cycles, 2 forms on closed 2-cycles etc. The number of axions arising due to the existence of a given form field is given by the number of closed cycles of the corresponding order; the relevant fields, however, are string theory dependent. All string theories contain the so-called ``model independent'' axion arising from compactification of the antisymmetric partner of the metric, $B_{\mu\nu}$, on closed 2-cycles. Generic string theory compactifications capable of realising realistic theories of high energy physics are highly complicated topologies, containing many hundreds of closed cycles (the source of the string landscape), and thus give rise to \emph{many axions}. 

The underlying symmetries require axions to possess a shift symmetry, $\theta\rightarrow \theta + 2 \pi$, and so their potential must be periodic. The Lagrangian for such an axion takes the following form:
\begin{equation}
\mathcal{L} = -\frac{f_a^2}{2}(\partial\theta)^2 - \Lambda^4 U(\theta)
\end{equation}
where $U(\theta)$ is some periodic potential. Bringing the kinetic term into canonical form we define the field $\phi = f_a \theta$, with Lagrangian:
\begin{equation}
\mathcal{L} = -\frac{1}{2}(\partial \phi)^2 - V(\phi)
\label{eqn:lagrangian}
\end{equation}
where $V(\phi)$ is again a periodic potential. Expanding the potential in powers of $\phi/f_a$, all the couplings of the field $\phi$ come suppressed by the scale $f_a$, and  from the quadratic term we find that the mass is given by:
\begin{equation}
m_a^2 = \frac{\Lambda^4}{f_a^2}
\end{equation}

The symmetry breaking scale, $f_a$ and the scale of the potential, $\Lambda$, are both determined separately for each axion, and depend on the action, $S$, due to non-perturbative physics on the corresponding cycle:
\begin{align}
f_a &\sim \frac{M_{pl}}{S} \\ \nonumber
\Lambda^4 &= \mu^4 e^{-S} \nonumber
\end{align}
where $M_{pl}$ is the reduced Planck mass: $M_{pl}^2=1/8\pi G$. Solving the strong $\mathcal{CP}$ problem requires $S\gtrsim 200$ \cite{witten2006,axiverse2009}, giving rise to stringy values of $f_a \approx 10^{16}\unit{GeV}$, and this should be roughly constant for all these axions. The exact value of $S$, however, scales with the area of the corresponding cycle (itself set by the scalar modulus partner of the axion), so that small variations in the area lead to exponential variations in the scale of the potential, and thus the axion mass. 

Hundreds of cycles of varying sizes therefore lead us to expect the appearance of at least some extremely light axions, given the following scenario. Axions generically get their masses lifted to high values at tree level, however the required lightness of the QCD axion necessary to solve the strong $\mathcal{CP}$ problem, therefore avoiding such liftings of the mass, implies that other axions too may survive as light and stable. If string theory solves the strong $\mathcal{CP}$ problem by giving us the QCD axion, then the axiverse appears as a natural source of light scalars for cosmology. Some authors have been able to explicitly construct realisations of this scenario \cite{acharya2010a,acharya2010b,higaki2011}, although the axiverse paradigm is expected to be much more general than these specific constructions.

The axion shift symmetry also enforces that couplings to fermions appear derivatively as $(\partial \phi)$, leading to factors of momentum at axion-fermion vertices. At low momentum this suppresses long range forces on fermions by factors of $k/f_a$ and string axions avoid fifth force constraints. In fact all axion couplings, including self couplings in the scalar potential, couplings to gauge fields like the photon due to higher dimensional operators, and topological couplings like the original $\theta\tilde{F}_{\mu\nu}F^{\mu\nu}$, come suppressed by this high scale. Therefore for cosmological purposes we will consider these axions to be completely decoupled, non-interacting massive scalar fields with potential:
\begin{equation}
V(\phi) = \frac{1}{2}m_a^2 \phi^2
\label{eqn:potential}
\end{equation}
Thus these axions are completely described by their mass, which we take to be a free parameter. We will consider dark matter axions with masses as low as $m\sim 10^{-32}\unit{eV}$ in the presence of additional CDM, neutrinos and a cosmological constant. Axion masses go as low as $m\lesssim 10^{-33}\unit{eV}$, at which point the axion behaves as quintessence, as in the scenario of \cite{frieman1995} described earlier. We will not consider quintessence axions.

Some interesting features of the cosmology of string axions and their relation to inflation, the production of gravitational waves and isocurvature perturbations were explored in \cite{fox2004,linde1991,hertzberg2008,mack2009a,mack2009b}. String axions may also effect astrophysical phenomena, for example through black hole super-radiance \cite{arvanitaki2010}.

Ultra-light scalar fields are known to share qualitative features and many degeneracies in their effects on cosmology with massive neutrinos and thus with many other cosmic ingredients \cite{amendola2005}. It is the aim of this paper to explore the effects of ultra-light string axions on the CMB and LSS, and thus these degeneracies, in detail.

\section{Ultra-light Axions vs Massive Neutrinos}
\label{neutrinos}

Light species of particles, such as massive neutrinos with $m_\nu \lesssim 1\unit{eV}$, can act as hot dark matter and suppress formation of large scale structure via free-streaming~\cite{bond1980}. On scales smaller than the free streaming scale, i.e. for wavenumbers $k>k_{FS}$,  the HDM cannot cluster.  This is determined by the temperature at which the species becomes non-relativistic, and therefore by the mass of the species. During matter or $\Lambda$ domination \cite{lesgourgues2006}:
\begin{align}
k_{FS}=0.82\frac{\sqrt{\Omega_\Lambda+\Omega_m(1+z)^3}}{(1+z)^2}\left( \frac{m}{1\unit{eV}} \right) h\unit{Mpc}^{-1}
\end{align}
If there is a fraction of matter, $f$, in such a non-clustering species then the overdensities in matter grow as $\delta \sim a^q$, with $q=1/4(-1+\sqrt{25-24f})$ for $k>k_{FS}$. This behaviour leads to the formation of ``steps'' in the matter power spectrum~\cite{amendola2005}. The size of these steps was first estimated in \cite{hu1998d} to be $\Delta P(k)/P(k) \approx -8 \tilde{f}_\nu$ ($\tilde{f}_\nu=\Omega_\nu/\Omega_m$). Fits for the steps can be found in \cite{eisenstein1999} and \cite{kiakotou2008}.

A qualitatively similar feature occurs in the presence of ultra-light scalar fields with $m_a \lesssim 10^{-18}\unit{eV}$, such as string axions, but the physics behind this process is quite different to the case of neutrinos or any other eV mass particles, such as the QCD axion~\cite{hu2000,amendola2005,axiverse2009,marsh2010}. 

A scalar field with the quadratic potential of Eq. \ref{eqn:potential}, decomposed into homogeneous and inhomogeneous components as $\phi(\vec{k},\tau)=\phi_0(\tau)+\phi_1(\vec{k},\tau)$ has the following equations of motion, to first order in cosmological perturbation theory about a homogeneous, flat FLRW background~\cite{bertschinger1995,hu1998b}:
\begin{align}
\ddot{\phi}_0 + 2\mathcal{H}\dot{\phi}_o + m^2 a^2 \phi_0 &= 0 \label{eqn:phi0} \\
\ddot{\phi}_1 +2\mathcal{H} \dot{\phi}_1 +(m^2 a^2 + k^2)\phi_1 &= -\frac{1}{2}\dot{\phi}_0\dot{h} \label{eqn:phi1}
\end{align}
where $a$ is the scale factor of the FLRW metric, overdots denote derivatives with respect to conformal time $\tau$, $\mathcal{H}=\dot{a}/a$ and $h$ is the scalar metric perturbation in conformal Newtonian gauge, as defined in \cite{bertschinger1995}. The density and pressure in the field are derived in the usual way, to first order, from the energy-momentum tensor, with these quantities again defined as in \cite{bertschinger1995}:
\begin{align}
\rho_a = &\frac{a^{-2}}{2}\dot{\phi}_0^2 + \frac{m_a^2}{2}\phi_0^2 \label{eqn:rhoa} \\
\delta\rho_a =& a^{-2}\dot{\phi}_0\dot{\phi}_1 + m_a^2 \phi_0 \phi_1 \label{eqn:deltarho}\\
P_a =& \frac{a^{-2}}{2}\dot{\phi}_0^2 - \frac{m_a^2}{2}\phi_0^2 \label{eqn:pa}\\
\delta P_a =& a^{-2}\dot{\phi}_0 \dot{\phi}_1 - m_a^2\phi_0 \phi_1 \label{eqn:deltap} \\
(\rho + P)\theta_a =&a^{-2}k^2 \dot{\phi}_0 \phi_1  
\end{align}

The equations of motion are oscillators, with oscillations beginning when the mass overcomes the Hubble friction in Eq.~\ref{eqn:phi0}. Using a WKB approximation we can solve Eqs.~\ref{eqn:phi0},~\ref{eqn:phi1} approximately and obtain the sound speed in the scalar field fluid perturbations, $c_s=\delta P/\delta\rho$, as an average over the period of oscillation~\cite{hu1998b,hu2000,marsh2010}:
\begin{align}
c_s^2 &= \frac{k^2}{4m_a^2 a^2}; \quad k<2m_a a \nonumber \\
c_s^2 &= 1; \quad k>2m_a a \nonumber \\
\label{eqn:cssquared}
\end{align}
On scales $k>2m_a a $ where the sound speed is $1$ the perturbations in the fluid are relativistic and overdensities will not grow, with the total overdensity in axions plus CDM scaling as $\delta\sim a^q$, just like the case of massive neutrinos. On scales below this the sound speed goes to zero and overdensities grow as $\delta\sim a$, just like pure dust CDM. This behaviour is related to the Compton wavelength of the ultra-light particles, and implements the idea FCDM~\cite{hu2000}.  

There emerges a new scale, $k_m$, analogous to the neutrino free streaming scale, for ultra-light scalars. Modes with $k>k_m$ enter the horizon whilst the sound speed is relativistic and will display a suppression of power, while modes with $k<k_m$ enter the horizon as the sound speed is decaying to zero and cluster as ordinary CDM. These considerations give:
\begin{align}
\frac{k_m}{H_0} &= (2\Omega_m)^{1/3} \left( \frac{m_a}{H_0} \right)^{1/3}; \quad k_m<k_{eq} \nonumber \\
\frac{k_m}{H_0} &= \left( \frac{4 \Omega_m}{1+z_{eq}} \right)^{1/4} \left( \frac{m_a}{H_0} \right)^{1/2}; \quad k_m>k_{eq} \nonumber \\
\label{eqn:km}
\end{align}
where $\Omega_m$ is the total fraction of the critical density in baryons, CDM, axions, and neutrinos (if they are non-relativistic at these scales), and $H_0$ is the Hubble scale today. We note that this definition of $k_m$ reproduces the same scaling with mass as the definition used in \cite{amendola2005} where $k_m$ is defined as the scalar field Jeans scale during matter domination evaluated at the redshift when scalar field oscillations begin. We further note that in our fits for the matter power spectrum made in subsequent sections we will only be considering $k_m<k_{eq}$, since massive neutrinos corresponding to the WMAP best fit values have $k_{FS}$ in this region. How much degeneracy there is between axions and neutrinos will clearly depend on exactly how close $k_m$ and $k_{FS}$ are and how sensitive a particular observable in a particular survey is to physics on these scales. 

In the fits we use $\bar{k}_m$ described in \cite{marsh2010}, where we add a bar to distinguish this fitted value, which fits the \emph{middle} of the step, from the value derived above, which fits the start. The two differ by an order of magnitude for our fiducial cosmologies, which reflects the scale over which the transition in axion clustering behaviour occurs. 
\begin{equation}
\bar{k}_m=A f_{ax}^{\alpha_1}(1+z)^{\alpha_2}(1-\Omega_\Lambda)^{\alpha_3}m^{1/3}
\end{equation} 
Ref. \cite{marsh2010} gave the values of the fitting parameters as approximately: $A= 1.25$, $\alpha_1 = -0.5$, $\alpha_2 = 0$, $\alpha_3 = 0.4$. The exponent $\alpha_3$ in these fits appears as $\Omega_m^{\alpha_3}$, and its fitted value of $\alpha_3=0.4$ is close to the expected value of $\alpha_3=1/3$. 

Finally we note that $k_m$ enters the horizon when $H\approx m_a$ ($\mathcal{H}=aH$), at exactly the same time when scalar field oscillations are expected to begin. This is again consistent with the definitions given in \cite{amendola2005}. However since Eq. \ref{eqn:cssquared} only holds once the fields have already begun oscillations we should take this as a warning that the expressions for $k_m$ given by Eq. \ref{eqn:km} will only be approximate.

Here we see the physical difference between suppression of structure by ultra-light scalars and the free-streaming of neutrinos. Free-streaming is related to a change in temperature causing the \emph{particles} to become non-relativistic when the temperature drops below the mass. Ultra-light scalars, if treated as particles, would still be relativistic today and their ``free-streaming'' scale would be larger than the horizon. However, ultra-light scalars form a \emph{condensate} (see, for example, \cite{sikivie2009} and references therein), and we treat them as a classical field. As such it is the sound speed of perturbations in this condensate, which depends only on the mass and scale factor, not the temperature, which determines whether overdensities can form. This leads to an interesting coincidence of scales: an ultra-light scalar with a mass in the range of $10^{-30}\unit{eV}$ will suppress structure on approximately the same scale as a neutrino with a mass $\mathcal{O}(10^{30})$ times greater, which will make up about 1\% of the total energy density. However, there is an extra parameter to consider for ultra-light scalars that comes from the production mechanism of their relic density and leads to an extra degree of freedom when considering their effects on the matter power spectrum.

The relic density of massive neutrinos is fixed by the mass of each neutrino species, assuming standard model interactions. For one massive neutrino this is approximately given by:
\begin{equation}
\Omega_\nu \approx \frac{m_\nu}{93.14 h^2 \unit{eV}}
\end{equation}
where $h$ is, and will from now on always be, the Hubble parameter defined as usual by $H_0 = 100 h \unit{km s}^{-1}\unit{Mpc}^{-1}$. We see that both $k_{FS}$ and $\Omega_\nu$ are fixed by the neutrino mass, $m_\nu$. However, as mentioned in Section \ref{introduction_ingredients}, for the case of sterile neutrinos this is not necessarily true. Sterile neutrinos are thermalised with active neutrinos via the mixing process, but the degree of thermalization depends strongly on the masses and mixing parameters \cite{melchiorri2009} and therefore, just like the case of $N_{\mathrm{eff, rel}}$, $N_{\mathrm{eff, mass}}$ can be given non-integer values to parameterise this. For thermalised standard neutrinos, since $\Omega_\nu$ is fixed by the mass, we have $k_{FS}=k_{FS}(\Omega_\nu)$. However for a sterile neutrino this is no longer true, since $\Omega_\nu=\Omega_\nu(m_\nu,N_{\mathrm{eff,mass}})$ \cite{hamann2010a}.

Axions have two contributions to their relic density. First of all there is standard thermal production due to axion couplings to the standard model. However, just like the self interaction terms in the potential, all of these couplings appear in the Lagrangian suppressed by powers of $f_a$. For large, stringy values of $f_a \sim 10^{16}\unit{GeV}$ these couplings are very small and the thermal relic density of ultra-light axions due to them is negligible.

There is also a second, non-thermal production known as the vacuum realignment mechanism. The axion arises from the spontaneous breaking of the Peccei-Quinn symmetry at the energy scale $f_a$. At this scale the parent scalar field acquires a vacuum expectation value and the Goldstone boson, which is the axion, acquires a \emph{random} initial value: the initial misalignment, $\phi_i$. Later, the field acquires its potential, $V(\phi)$, due to non-perturbative physics and once the mass of this potential overcomes Hubble friction, $H(z_{osc})\approx m_a$, the field will roll and generate a relic density that depends on the initial misalignment.  

\begin{figure*}
\centering
$\begin{array}{@{\hspace{-0.2in}}l@{\hspace{-2.3in}}l}
\includegraphics[scale=0.45,trim=18mm 0mm 10mm 0mm,clip]{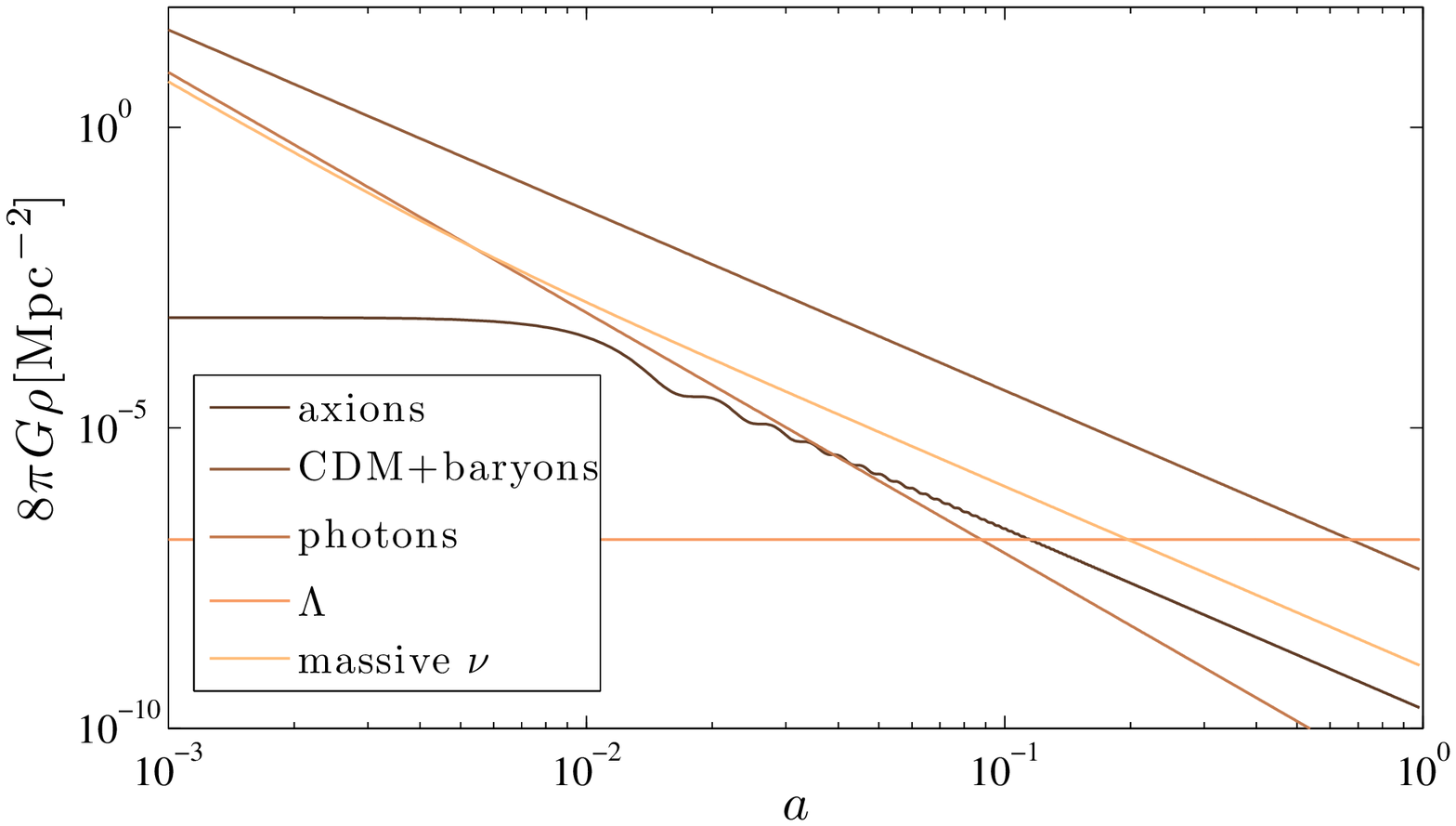}&
\includegraphics[scale=0.45,trim=18mm 0mm 10mm 0mm,clip]{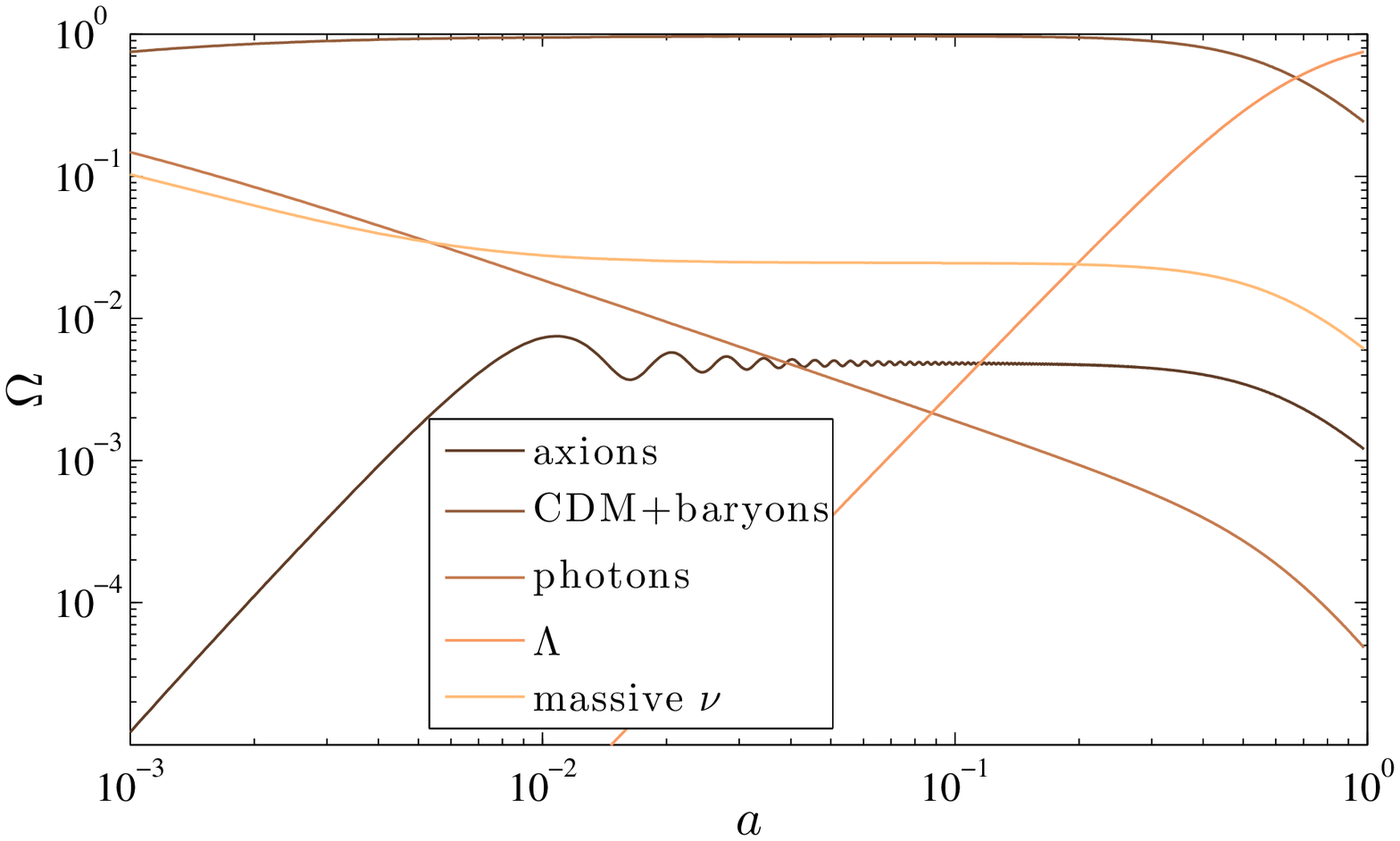}\\[0.0in] 
\end{array}$
\caption{Ultra-light axion and massive neutrino evolution in the background. $m_a=10^{-30}\unit{eV}$, $\Omega_a=0.1\Omega_d$, $N_{\mathrm{eff,mass}}=3$, $m_\nu=0.1\unit{eV}$. Left panel: evolution of the energy densities in massive neutrinos and axions compared to vanilla cosmological components. Right Panel: Contributions to the critical density, $\Omega_i\equiv \rho_i/3H^2$.}
\label{fig:rho_and_omega}
\end{figure*}

Until $z_{osc}$ the axion contributes negligibly to the energy density as a cosmological constant, in contrast to a massive neutrino, which scales as radiation before becoming non-relativistic. The effects on the redshift of equality by axions and neutrinos are thus not the same. Axions reduce the amount of matter only and do not affect the expansion rate while frozen; once rolling they have but a transitory effect on the background expansion away from $\Lambda$CDM as they go through their first few oscillations \cite{marsh2010}. Massive neutrinos, or extra relativistic species, in our parameterisation reduce $\Omega_\Lambda$ compared to redshift zero, but also increase the amount of radiation at early times. This has a non-negligible effect on the expansion rate and leads to a  markedly different effect in small scale CMB anisotropies between axions and neutrinos, as we discuss in Section~\ref{cmb}. Fig. \ref{fig:rho_and_omega} shows the evolution of the energy density in various components.

Using that $z_{eq}=2.5\times 10^4 \Omega_m h^2 \Theta_{2.7}^{-4}$, where $\Theta_{2.7}=T_{CMB}/2.7$, and the neutrino sector consists of three standard massless neutrinos, the relic density produced by this mechanism is approximately given by:
\begin{align}
\Omega_a &= 8.4 \times 10^{-5} h^{-3/2} \Theta_{2.7}^3 \left( \frac{m}{H_0} \right)^{1/2} \phi_i^2; \quad z_{osc}>z_{eq} \label{eqn:omegarad} \\
\Omega_a &= \frac{1}{6} \Omega_m \phi_i^2; \quad z_{osc}<z_{eq} \label{eqn:omegamat}
\end{align}
where $\phi$ is dimensionlessly given in Planck units.

For a quadratic potential, where we have explicitly broken the axion shift symmetry and are strictly working with generalised ultra-light scalars, we are essentially free to choose $\phi_i$ to give us the desired relic density, be it large or small, for any axion mass. Therefore for axions the scale for suppression of power and the relic density are separately under control via the two parameters $m_a$ and $\phi_i$, in contrast to standard massive neutrinos where both are fixed by the mass, $m_\nu$. It is, however, pertinent to consider questions of fine tuning for this production mechanism. We direct the reader to the discussions of \cite{mack2009a,mack2009b,marsh2010,tegmark2006}. 

Another feature of vacuum realignment affecting the axion relic density comes from the periodicity of the axion potential. The periodicity, if the shift symmetry remains unbroken, leads to the axion field having a maximum value given by $\phi_{max}=\pi \frac{f_a}{M_{pl}}$. Therefore it is clear that for axions below a certain mass it is impossible, barring anharmonic effects in the potential, to produce $\mathcal{O}(1)$ values for $\Omega_a$, which leads to the existence of what the authors of \cite{axiverse2009} call ``the anthropic window''. However, in the spirit of cosmological parameterisation, this need not worry us. If our model requires a larger amount of axion energy density at a certain mass than we can produce with $\phi_i\leq\phi_{max}$ then we can view this in the same way as $N_{\mathrm{eff}}$: it may be telling us that there are many species of axion in that mass range, where the masses cannot be resolved. This may, however, require additional fine tuning within the axiverse. For the fiducial models considered for forecasts in this work, however, we do not saturate this bound, and so the shift symmetry is preserved.

Standard model neutrinos come in three species, and each species should be massive, with some hierarchy and degeneracy structure between them. The cosmological detection of this degeneracy using weak lensing was discussed in \cite{bernardis2009}. In the axiverse scenario we have multiple species of axion, with their own mass splittings. Naively, then, because of the qualitative similarity in their effects on cosmology, we may expect to account for any discrepancy between terrestrial measurements of $N_{\nu}$, $m_\nu$ and mass splittings with the values determined by cosmology via the introduction of ultra-light scalars. In contrast, the possible existence of sterile neutrinos and other relativistic relics may obscure the possible cosmological effects of axions and close this observational window on them. It is one of the principle aims of this paper to go some way towards addressing these potential degeneracies, and indeed we expect many of them to be broken by considering multiple cosmological probes in the CMB and LSS, in the same way as degeneracies between neutrinos, dark energy and initial conditions can be broken in this way \cite{hu1998d,hannestad2006,joudaki2011}. However, due to complications in forecasting for the effects of a varying axion mass and of the effect of a neutrino hierarchy splitting, we will leave the analysis of this particular degeneracy for a future work, and here focus in our forecasts purely on the density for a single species of axion, and on degenerate massive neutrinos.

We summarise in Table \ref{tab:scales} the relevant scales of $z_{osc}$, $k_m$, and $\bar{k}_m$ for the axions used in our fiducial cosmologies. We also quote $k_{eq}(f_{ax}=0)$, $k_{eq}(f_{ax})$ (where $f_{ax}=\frac{\Omega_a}{\Omega_d}$), and $k_{FS}(m_\nu,z=0)$ for comparison. Note that although we always have $k_m<k_{eq}$, this is not always the case for $\bar{k}_m$. We also note that one's definition of $z_{osc}$ is somewhat ambiguous: does one define it from when $m_a=H$ or when $m_a=3H$, or somewhere in between; when slow roll is broken, or when the oscillations have settled down to CDM behaviour? This leads to an $\mathcal{O}(1)$ multiplicative factor of uncertainty. In particular, this makes $z_{osc}$ with $m_a=10^{-29}\unit{eV}$ potentially very close to $z_{rec}\sim 1100$.

\begin{table}[dtp]
\begin{center}

\begin{tabular}{|c|c|c|c|}
\hline
$m_a$ (eV) & $k_m (h $Mpc$^{-1})$& $\bar{k}_m (h $Mpc$^{-1})$ & $z_{osc}$ \\
\hline
$10^{-29}$ & 0.0058 & 0.0575 & 350\\
$10^{-30}$ & 0.0027 & 0.0267 & 74\\
$10^{-31}$ & 0.0012 & 0.0124 & 15\\
$10^{-32}$ & 0.0006 & 0.0057 & 2.4\\
\hline
\end{tabular}\\
$k_{eq}(f_{ax}=0)=0.0136 h\unit{Mpc}^{-1}$\\
$k_{eq}(f_{ax}=0.01)=0.0135 h\unit{Mpc}^{-1}$ \\
$k_{FS}(m_\nu=0.055\unit{eV},z=0)=0.0451h\unit{Mpc}^{-1}$
\end{center}
\caption{Relevant scales for our fiducial cosmologies with $f_{ax}=0.01$. $k_m$ is the scale at which structure suppression begins, given by Eq. \ref{eqn:km}. $\bar{k}_m$ is the location of the middle of the induced feature in $P(k)$, fit for in \cite{marsh2010}. $z_{osc}$ is the redshift at which axion oscillations begin, which has an $\mathcal{O}(1)$ multiplicative uncertainty.}
\label{tab:scales}
\end{table}

\section{The Axiverse and Cosmological Observables}
\label{observations}

\begin{figure}
\includegraphics[scale=0.5,trim=23mm 5mm 10mm 0mm,clip]{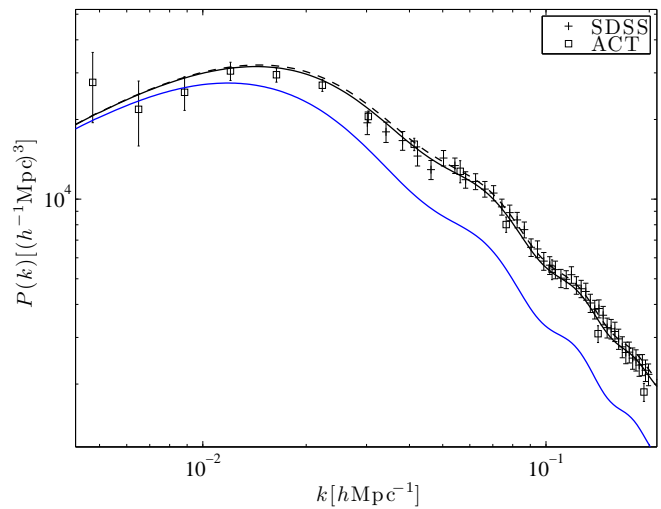}
\caption{The matter power spectrum for three cosmologies shown with current measurements from SDSS \cite{reid2010}, and ACT \cite{hlozek2011}. We show first the WMAP7 cosmology (dashed black line). We also show two axion cosmologies, both with $m_a=10^{-29}\unit{eV}$: $f_{ax}=0.1$ (solid blue line), and $f_{ax}=0.01$ (solid black line), with all other parameters held fixed at their WMAP7 values. Both axion cosmologies have only a small effect on the CMB power spectrum, but are clearly distinguished in their effect on $P(k)$, with $f_{ax}=0.1$ clearly ruled out by the data.}
\label{fig:pk_data}
\end{figure}

As discussed in Section~\ref{neutrinos}, ultra-light axions give rise to steps in the matter power spectrum, $P(k)$. Fig. \ref{fig:pk_data} shows this effect on the cosmology of WMAP7 \cite{7yearWMAP}, with the introduction of a single axion species with fraction $f_{ax}$. Large axion fractions, disallowing variation of other parameters, can easily be ruled out at current sensitivity, while a small fraction of around 1\% is indistinguishable from $\Lambda$CDM using the power spectrum of SDSS alone \cite{reid2010}, c.f. Fig. 1 of \cite{hu1998d}. Fig. \ref{fig:pk_data} also shows power spectrum constraints coming from the ACT measurement of the primordial power spectrum \cite{hlozek2011}. This appears to be able to rule out a 10\% fraction in axions easily using the CMB alone, which we will see is not the case for Planck. The reason being that these data points are evolved from the primordial power \emph{assuming pure CDM} in the transfer function. This is just one example, of which we will see others later, of the way in which we might naively misinterpret data if we do not assume the correct underlying cosmology. We will see that the small fractions of axions in our fiducial models, while still indistinguishable from $\Lambda$CDM with a single observable at a single redshift, can be distinguished using redshift information and/or a combination of observables.

In this section we discuss in detail the theoretical effects of ultra-light axions on the various cosmological observables. The effects are explored both analytically, using the fits of \cite{marsh2010,eisenstein1998}, and through numerical solution of the Boltzmann equations obtained from a modified version of the publicly available code CAMB \cite{lewis2000,lewis2002,camb}. Our modification introduces a module to deal with scalar fields having a quadratic potential with mass large compared to the Hubble rate, the bulk of which involves accurately fixing the initial conditions and background evolution in the presence of rapid oscillations, and integrating such oscillations accurately \footnote{For more details, or a copy of the code, please contact us via e-mail.}.

We exactly numerically solve the evolution of the axion field, $\phi$, the difficulty of which stops us exploring the region of parameter space with $m_a \gtrsim 10^{-28}\unit{eV}$, suggested by \cite{axiverse2009} to be the most interesting region to look for unique step-like features in the power spectrum with a high precision galaxy survey or 21cm tomography survey. We are also limited to studying a single axion field, however our results will show that in fact, since constraints from some observables are mass independent, this is not a limitation. Our technique makes no use of the approximate treatments of axion sound speed and averaging used in the analysis of \cite{amendola2005}. In addition, the mass range that we study is the one found in \cite{amendola2005} to have the most tightly constrained axion fraction, but also the range in which the approximations used are least sound. Future observations will bound this regime even more tightly; making reliable predictions for high precision measurements in this important regime requires an exact treatment such as ours.

Throughout this section we will use our physical intuition about the suppression of power caused by ultra-light axions, and the similarities and differences with respect to neutrino free-streaming, to try and understand our numerical results. Where possible, we will be guided by analytic fits, but stress that these are meant for qualitative purposes only, and have some limited applicability, which we discuss. Analytic fits are not used in our forecasts. We emphasise that the figures and discussion of parameter variation in this section are meant only for illustrative purposes, and are not meant in any way as parameter estimation from existing data, nor do they necessarily reflect the fiducial models of our forecasts. A Markov-Chain Monte-Carlo analysis for parameter estimation in this model will be the subject of a future work.

\subsection{The Matter Power Spectrum}
\label{matterpower}

In \cite{marsh2010} we derived fits, $T_{ax}(k,z,f_{ax})$, for the shape of the steps in $P(k)$ in a flat universe containing radiation, axions, CDM, and a cosmological constant, $\Lambda$, but no baryons:
\begin{align}
\tilde{f}_d T_d (k,z,\tilde{f}_{ax})&=\tilde{f}_c T_c(k,\tilde{f}_{ax})+\tilde{f}_{ax}T_a(k,z,\tilde{f}_{ax})\nonumber \\
			    &=  \tilde{f}_d T_{ax}(k,z,\tilde{f}_{ax})T_c(k,\tilde{f}_{ax}=0)
\end{align}
Here, and throughout this paper, $f_{i}=\Omega_i/\Omega_d$, $\tilde{f}_i=\Omega_i/\Omega_m$ so that $\tilde{f}_i=f_i \tilde{f}_d$. In \cite{marsh2010}, these quantities were equal since $\tilde{f}_d=1$ in the absence of baryons. The difference between $f_{ax}$, $\tilde{f}_{ax}$ is important in the functional form of $T_{ax}$, as we will see below. We explicitly show the redshift dependence of $T_d$ arising from $T_{ax}$, which corresponds to scale dependent growth. Whenever we drop redshift dependence, it is assumed that $z=0$. 

The matter power spectrum is related to the transfer function by: $P(k,z)=A(k)T_m^2(k)D_1^2(z)$, where $D_1(z)$ is the growing mode, given for example in \cite{eisenstein1999}, and $A(k)$ is the primordial power. Therefore, the step in the matter power spectrum in the model of \cite{marsh2010} was given by:
\begin{equation}
T^2_{ax}(k,z,f_{ax})=\frac{P(k,z,f_{ax})}{P(k,f_{ax}=0)}
\end{equation}

In our numerical studies using CAMB the cosmology contains, in addition to CDM, axions and $\Lambda$ considered in \cite{marsh2010}, the other standard ingredients of baryons, and their coupling to photons, massless and massive neutrinos. In the presence of baryons we model the full matter transfer function according to \cite{eisenstein1998} as:
\begin{equation}
T_m(k)=\tilde{f}_d T_d(k)+\tilde{f}_b T_b(k)
\end{equation}
$T_d(k)$ is the total dark matter transfer function, including CDM, axions and massive neutrinos, if present, and $T_b(k)$ is the baryon transfer function. The fitted baryon transfer function contains the gravitational effects of the coupling to dark matter through its dependence on the matter fraction $\Omega_m=\Omega_d+\Omega_b$, the sound horizon $s$, the drag epoch $z_d$, the epoch of equality $z_{eq}$, the scale of equality $k_{eq}$, and the Silk damping scale $k_{Silk}$. There is also a dependence on these scales incorporated into the dark matter transfer function. In the case of ultra-light axions that do not begin their oscillations until the matter dominated era, these scales should all be altered to account for the change in matter content during these epochs. For example, $z_{eq}\rightarrow \tilde{f}_{c+b}z_{eq}$. Since the gravitational effect of the dark matter has thus already been accounted for in the baryon transfer function, the step feature modelled by $T_{ax}(k,z,f_{ax})$ should only multiply the DM transfer function, but with the weighting for axion effects coming in as $\tilde{f}_{ax}=\Omega_a/\Omega_m$ that is $T_d(k)\rightarrow T_d(k,z,\tilde{f}_{ax})$. The total matter transfer function is thus given by:
\begin{equation}
T_m(k,z,\tilde{f}_{ax})=\tilde{f}_d T_{ax}(k,z,\tilde{f}_{ax})T_c(k,\tilde{f}_{ax}=0)+\tilde{f}_b T_b(k,\tilde{f}_{ax})
\label{eqn:tmfit}
\end{equation}
where $T_c(k,\tilde{f}_{ax})$ and $T_b(k,\tilde{f}_{ax})$ are fit by the formulae of \cite{eisenstein1998}, with relevant scales modified by the presence of the ultra-light component. The distinction between $f_{ax}$ and $\tilde{f}_{ax}$ is especially important in the explicit form of $T_{ax}$ from \cite{marsh2010}.

Therefore the step in the power spectrum caused by  an axion component in the presence of baryons is given by:
\begin{equation}
\tilde{T}^2_{ax}(k,z,\tilde{f}_{ax})=\frac{T^2_m(k,z,\tilde{f}_{ax})}{T^2_m(k,\tilde{f}_{ax}=0)}
\label{eqn:newtfit}
\end{equation}
where $T_m(k,z,f_{ax})$ is given by Eq. \ref{eqn:tmfit}. Note that this fit is related to the fit of \cite{kiakotou2008} used to investigate the effects of massive neutrinos by $\Delta P(k)/P(k) = \tilde{T}_{ax}^2(k,\tilde{f}_{ax})-1$.
In particular, the step size is defined by:
\begin{equation}
\tilde{S}(\tilde{f}_{ax})=\lim_{k \to \infty}\tilde{T}^2_{ax}(k,\tilde{f}_{ax}) \nonumber
\end{equation}
Given that the baryon transfer function goes to zero faster than the CDM transfer function as $k$ goes to infinity, we have:
\begin{equation}
\tilde{S}(\tilde{f}_{ax})=S(\tilde{f}_{ax})
\end{equation}
where $S$ is defined in \cite{marsh2010}, which produces a \emph{smaller} step than the case with no baryons, $S(\tilde{f}_{ax})>S(f_{ax})$. This is caused by axions making up a smaller fraction of the total matter than of the dark matter alone, i.e. $\tilde{f}_{ax}<f_{ax}$.

The  ``naive'' fit: $\tilde{T}_{ax}(k,\tilde{f}_{ax})=T_{ax}(k,\tilde{f}_{ax})$, corresponds to axions suppressing growth on CDM and baryons evenly, with no account made for axion effects on the sound horizon, drag epoch etc. This, as expected, reproduces the small scale limit. The modified fit for $\tilde{T}_{ax}(k,\tilde{f}_{ax})$ incorporates changes to the sound horizon, Silk damping scale and drag epoch using the fits of \cite{eisenstein1998} and thus qualitatively captures the deviations from the smooth fit due to distortions of the Baryon Acoustic Oscillations (BAO), which are seen in the full numerical solution. However, the fits presented here end up overestimating the total amount of power suppression by a few percent.

A step in the power spectrum corresponds to a change in the ratio of small to large scale power, which can naively be mimicked by changes in other cosmological parameters, such as the tilt of the primordial power spectrum. If measurements at large scales are poor, the effect can also be mimicked by adding more CDM, which shifts the power spectrum over to larger $k$. Isolating the unique effect of a structure suppressing species requires precise observations at the relevant scale, $k_{FS}$ or $k_m$ \cite{hu1998d}. We are considering axion species varying in mass over many orders of magnitude, so have a correspondingly large variation of the scale $k_m$. Power spectrum measurements have varying precision over this range of $k_m$, and so we expect different constraints on the axion fraction, and possibly different degeneracies with other cosmological parameters, for the axions of different masses, if $k_m$ for the different species falls into regions of different accuracy in the survey. In particular, we should expect stronger constraints from galaxy redshift surveys alone on heavier axions with larger $k_m$. However, since all the axions we consider have $k_m<k_{eq}$, where survey accuracy is at its lowest, this effect should not be significant.

\subsection{Baryon Acoustic Oscillations}
\label{BAO}


The theory behind Baryon Acoustic Oscillations (BAO) and their effect on the matter power spectrum has been known since 1998 in the work of Eisenstein and Hu \cite{eisenstein1998}, and they are an important cosmological tool used in the distance ladder (see for example \cite{percival2007,percival2009,2001MNRAS.327.1297P,2003ApJ...598..720S,2005ApJ...633..560E,2010MNRAS.401.2148P,beutler2011}, and most recently \cite{blake2011}, and for a review see \cite{bassett2009}). Measuring the BAO to high precision is thus a key goal in modern cosmology. Here we briefly discuss a method of extracting, and hence working definition of, BAO from the matter transfer function.

We define the linear BAO as in e.g. \cite{percival2007}:
\begin{equation}
B_{lin}=\frac{T^2_{m,\mathrm{full}}(k)}{T^2_{m,\mathrm{no\ osc}}(k)}
\label{eqn:blin}
\end{equation}
where $T_{m,\mathrm{full}}(k)$ is the matter transfer function for a certain cosmology, either numerical or analytical, and $T_{m,\mathrm{no\ osc}}(k)$ is defined as the oscillation free, smooth transfer function taken as an $n-$node cubic spline of $T_{m,\mathrm{full}}(k)$ at points $k_i$, $i=1,\ldots, n$ chosen empirically to get the best smooth fit.

There is a small distortion of $\tilde{T}_{ax}(k,\tilde{f}_{ax})$ in a cosmology with baryons, away from its smooth form in the region of the BAO. We now aim to give some analytic understanding of the reason for this, and therefore predict how large we can expect any BAO distortions due to light axions to be.

Since the gravitational effect of the dark matter has thus already been accounted for in the baryon transfer function, the step feature modelled by $T_{ax}(k,z,f_{ax})$ should only multiply the DM transfer function, but with the weighting for axion effects coming in as $\tilde{f}_{ax}=\Omega_a/\Omega_m$ that is $T_d(k)\rightarrow T_d(k,z,\tilde{f}_{ax})$.  We fit $T_c(k,\tilde{f}_{ax})$ and $T_b(k,\tilde{f}_{ax})$ using the formulae of \cite{eisenstein1998}, with relevant scales modified by the presence of the ultra-light component. We see that while $T_{ax}$ is smooth and should thus be captured by the spline in $B_{lin}$, we have also introduced a dependence on axion fraction into the baryon transfer function, which is oscillatory.

In Eqn. 21 of \cite{eisenstein1998} it is clear that the amplitude of the oscillations in the baryon transfer function has a detailed dependence on all the cosmological scales: the Silk damping scale, the scale of equality, the redshift of the drag epoch, and the sound horizon at the drag epoch, and all of these scales depend on the matter content at the time when they are relevant, and at all such scales the lightest axions were frozen and not contributing as matter. The dependence of the sound horizon on the redshifts of drag and equality further implies that varying these redshifts by including an exotic species will cause a variation in the period of the BAO, as we can see again from Eqn. 21 of \cite{eisenstein1998}, or from the fit used in \cite{percival2007}. 

If we hold the total amount of dark mater fixed and introduce ultra-light axions with $z_{osc}<z_{eq}$ then all these scales are shifted relative to where they would be if the dark matter were pure CDM. The shift is simple to compute: we simply alter the matter content in the equations for calculating these scales by a factor $(1-\tilde{f}_{ax})$ to account for the frozen axions. The distortions in this case can be easily understood from \cite{eisenstein1998}. We effectively change $\Omega_m(z)$ during specific epochs ($\Omega_0$ in the notation of \cite{eisenstein1998}) while keeping the fraction $\Omega_b/\Omega_m$ (at $z=0$) constant. In a cosmology with massive neutrinos we must further shift these quantities by increasing the density in relativistic species at the relevant epochs appropriately for the mass of the neutrino species: a more complicated effect. Note that such distortions in the case of massive neutrinos are not modelled in either \cite{eisenstein1999} or \cite{kiakotou2008}, where baryon oscillations are not present: it is only the scale dependent growth due to massive neutrinos that is considered there (see Section \ref{growthrate}).

\begin{figure}
\centering
$\begin{array}{@{\hspace{-0.25in}}l@{\hspace{-1.5in}}l}
\includegraphics[scale=0.4,trim=18mm 10mm 10mm 0mm,clip]{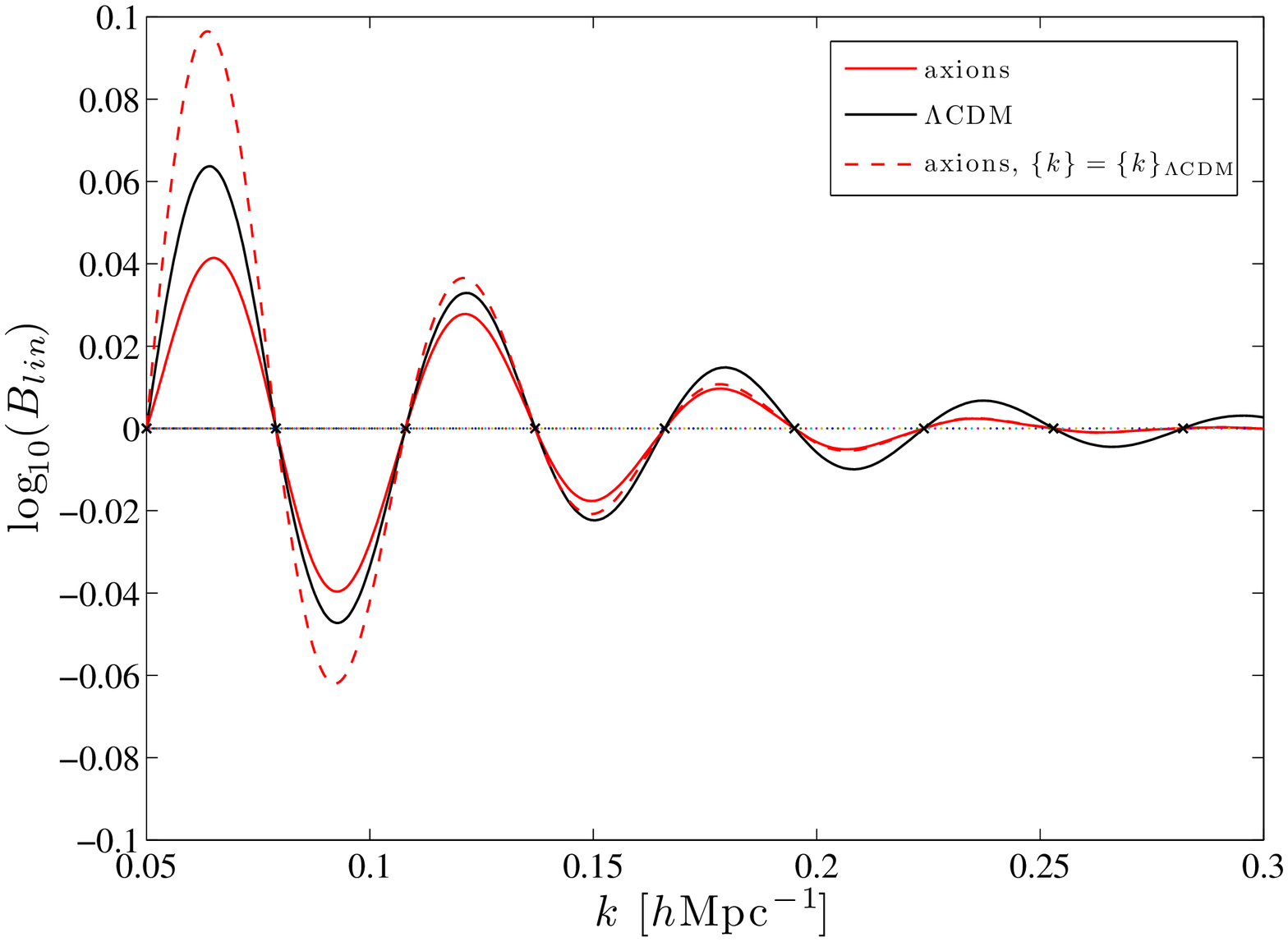}
\end{array}$
\caption{The BAO (at $z=0$) for two models: a fiducial $\Lambda$CDM model, and a model with axions having $m_a=10^{-30}\unit{eV}$, $f_{ax}=0.1$. We also show the BAO for the same axion model, but where the smooth transfer function is splined assuming $\Lambda$CDM.}
\label{fig:bao1}
\end{figure}

In Fig.~\ref{fig:bao1} we show the BAO calculated numerically including the effects of ultra-light axions with $m=10^{-30}\unit{eV}$, such that $k_m < k_{eq}$,  keeping $\Omega_d$ constant, and fitting a smooth $T_{\mathrm{no\ osc}}(k)$ using a cubic spline. We see an overall \emph{suppression} of BAO amplitude caused by the presence of axions, despite an increased ratio $\Omega_b/\Omega_d$ in the radiation dominated era.

\begin{figure}
\centering
$\begin{array}{@{\hspace{-0.25in}}l@{\hspace{-1.5in}}l}
\includegraphics[scale=0.45,trim=18mm 10mm 10mm 0mm,clip]{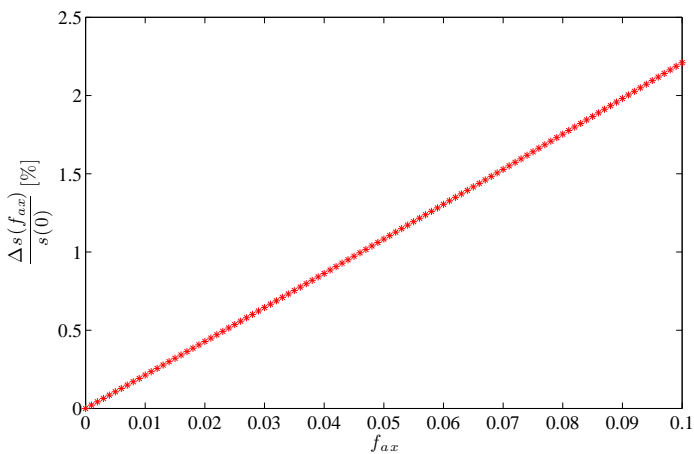}
\end{array}$
\caption{Change in the sound horizon at recombination as function of axion fraction as a percentage of its zero axion value, calculated using the formulae of \cite{eisenstein1998}.}
\label{fig:sound_horizon}
\end{figure}

Some model dependence enters in our definition of $T_{\mathrm{no\ osc}}(k)$. For our $\Lambda$CDM model, we find empirically the best $k$ points for our cubic spline are $k=0.001$ and $0.029\leq  k \leq 0.369$ with $\Delta k=0.05$\footnote{This is slightly different from the values chosen in \cite{percival2009}.}. The step in the transfer function caused by axions is a smooth feature, and is best captured by introducing two extra points into the spline at $k = 0.0081,0.02$, where the first is our estimate for $k_m$ in this cosmology. The frequency of the BAO is fixed by the choice of the set $\{0.029\leq  k \leq 0.369,\Delta k=0.05\}$, since the spline anchors $B_{lin}=1$ at these values, therefore we cannot see any change in the frequency in the figure. That the same set fits $T_{\mathrm{no\ osc}}(k)$ (by eye) both with and without axions tells us that any change in frequency is small. To quantify this, we note that the frequency of the BAO is set by the sound horizon as $\sin (s k)$ \cite{percival2007}. In Fig. \ref{fig:sound_horizon} we plot the change in the sound horizon calculated using the formulae of \cite{eisenstein1998} as a function of $f_{ax}$ as a percentage of its zero axion value. Axions always increase the size of the sound horizon, but even for large fractions the change is only by a few percent. The increase is linear, a fact not at all obvious from the relevant formulae.

We may ask whether the BAO can be biased by our choice of $T_{\mathrm{no\ osc}}(k)$, effectively the underlying cosmology that we assume. In Fig.~\ref{fig:bao1} we show an example of such bias by choosing to instead fit a smooth transfer function using the $k$ points necessary for a $\Lambda$CDM cosmology, with no extra points to fit the step. This leads to an increase in the apparent BAO amplitude on large scales in the first few oscillations.

There is, according to our fits, a significant degeneracy between $f_{ax}$ and $\Omega_c h^2$. We can introduce more CDM to fix the epoch of equality while ultra-light axions are present. Fixing a flat universe, this will reduce $\Omega_\Lambda$ by some small amount, the effect on the power spectrum being only through the normalisation \cite{eisenstein1998}, and invisible in ratios such as the BAO and $\tilde{T}_{ax}$. Restoring equality also restores the Silk damping scale and the drag epoch to their axion-free values, and this restoring effect dwarfs any small changes to the BAO through the alteration of the matter fractions $\tilde{f}_d$, $\tilde{f}_b$. In this case the BAO distortions can be removed, and the overall suppression of power due to axions, $\tilde{T}_{ax}$, is also reduced as the increase in CDM shifts the power spectrum over to larger $k$. Avoiding such an alteration to the BAO will, however, have obvious effects on the CMB through changing the entropy per baryon, and will need further compensation for example in fitting the Hubble expansion rate by lowering the DE equation of state, $w$. We discuss some of these issues in Sections~\ref{weaklensing_theory} and \ref{cmb}.

Real BAO measurements depend on redshift, can measure the expansion rate as a function of $z$, which can break some of the simple degeneracies discussed here. In addition, their measurement in models including axions will therefore be complicated by the scale dependent growth introduced by the $z$ dependence of $T_{ax}$, which we now discuss.



\subsection{The Growth Rate}
\label{growthrate}

The growth rate is another useful cosmological observable, which has been measured by \cite{peacock2001measurement,ross3812df,guzzo2008test,2011MNRAS.415.2876B}, and is used particularly in studies of modified gravity. It is defined as:
\begin{equation}
f = \frac{d \ln \delta}{d \ln a}= \frac{\dot{\delta}}{\mathcal{H}\delta}
\label{eqn:growthdef}
\end{equation}
In standard $\Lambda$CDM it is known to be approximately scale invariant and behave with redshift as $f= \Omega_m(z)^{\gamma_g}$ (this is a useful approximation, but see for example \cite{giovannini2011} for a recent discussion). We expect the growth rate for cosmologies including axions or massive neutrinos to pick up some additional scale dependence related to the appearance of steps in the matter power spectrum \footnote{For a discussion on scale-dependent growth, see \cite{acquaviva2010} and references therein.}. 

\begin{figure}
\includegraphics[scale=0.4,trim=25mm 5mm 10mm 0mm,clip]{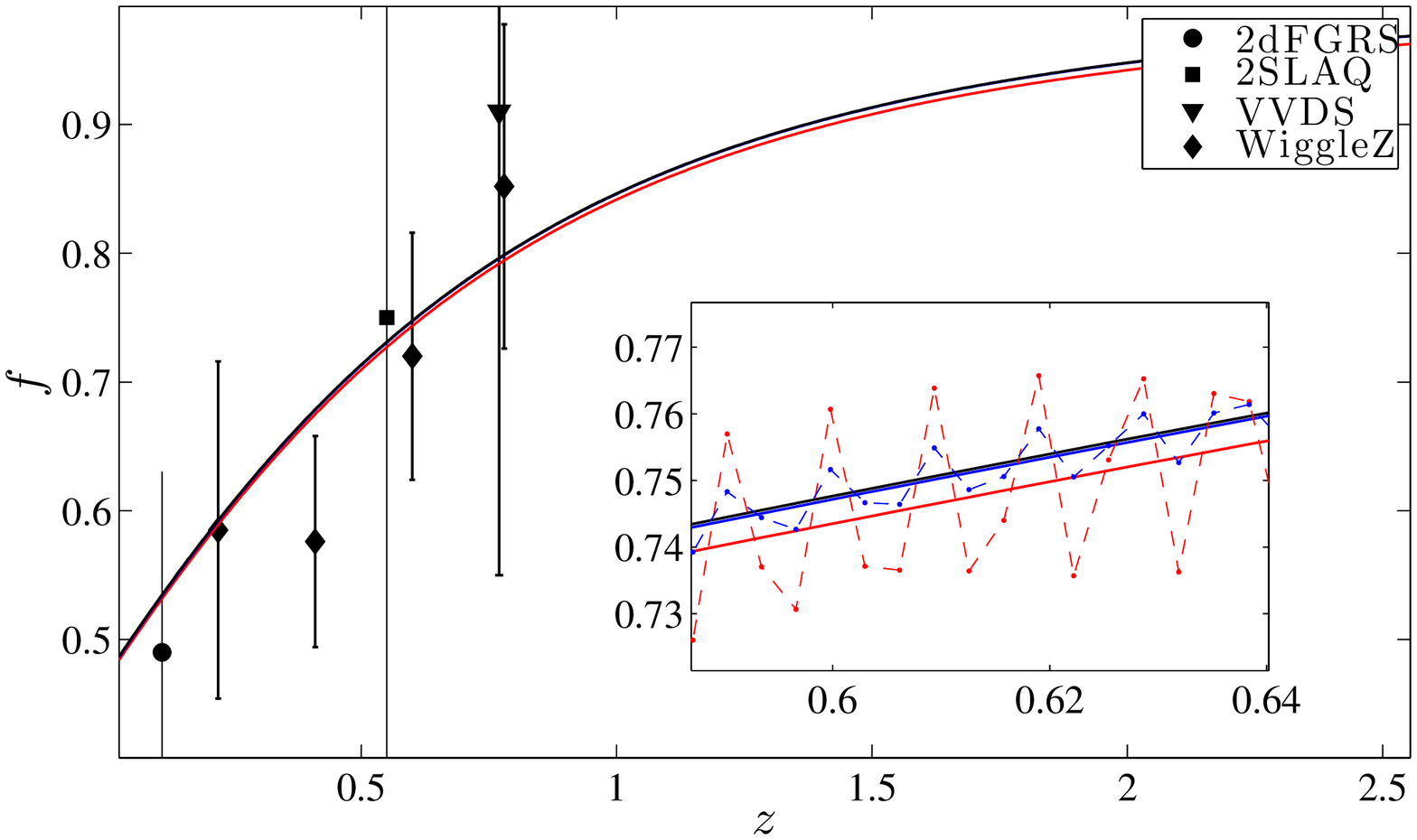}
\caption{The (reduced) growth rate (Eq.~\ref{eqn:reduced_growth}), $f(z)$, for $m_a=10^{-30}\unit{eV}$, $f_{ax}=0.01$, with current measurements of \cite{peacock2001measurement,ross3812df,guzzo2008test,2011MNRAS.415.2876B}. Red: $f_{(c+b)}(k=k_{max})$; blue: $f_{(c+b)}(k=k_{min})$; black: $\Lambda$CDM. Insert: zoom in to an arbitrary region of $z$ showing oscillations in the true growth rate (dashed lines) about the reduced growth rate. The reduced growth rate is seen to be a good qualitative tracer of the average, following the shape in $z$ and demonstrating scale dependence in its amplitude.}
\label{fig:growth1}
\end{figure}

In Fig.~\ref{fig:growth1} we plot the growth rate as function of $z$ for a standard $\Lambda$CDM cosmology, and for a cosmology with a fraction of axions $f_{ax}=0.01$. The growth rate with axions is plotted at two different $k$-values: the highest and lowest from $P(k)$ being $k_{min}\approx 10^{-5}h\unit{Mpc}^{-1}$, $k_{max}\approx 0.3h\unit{Mpc}^{-1}$. The true growth rate with axions (insert of Fig.~\ref{fig:growth1}) is seen to contain rapid oscillations that obscure the details of any step in $f(k)$ at a given redshift. These oscillations are not observable, and should be averaged over to be consistent with the interpretation of scalar fields behaving as dark matter in the background, which is also only true on average as seen from the WKB solution to Eq. \ref{eqn:phi0}. To account for this easily we present an approximation scheme that is valid for both axion and massive neutrino effects on the growth rate when the fraction in these species is small, as is true for all of our fiducial models.

We split the density into two pieces: $\rho_{c+b}$ in the dominant CDM and baryon components, and $\rho_{a+\nu}$ in the sub-dominant axion and massive neutrino components, and do the same for the perturbations. Using that $\rho_{a+\nu}$ is of order a few percent of $\rho_{c+b}$ we expand the expression for $f$ and drop terms of order $\rho_{a+\nu}^2$ and $\rho_{a+\nu}\delta\rho$ as being second order. Next we note that the overdensities will also be predominantly made up of CDM and baryons and so further expand in powers of $\delta\rho_{a+\nu}/\delta\rho_{c+b}$ and take terms of order $(\delta\rho_{a+\nu}/\delta\rho_{c+b})^2$, $\rho_{a+\nu}(\delta\rho_{a+\nu}/\delta\rho_{c+b})$ and $\delta\rho_{a+\nu}/\delta\rho_{c+b}^2$ as second order. The resulting expression for the growth rate is given by:
\begin{equation}
f = f_{(c+b)} - \left( \frac{\delta\rho_{a+\nu}}{\delta\rho_{c+b}} \right) (3+f_{(c+b)}) + \frac{1}{\mathcal{H}}\frac{d}{d \tau}\left(  \frac{\delta\rho_{a+\nu}}{\delta\rho_{c+b}} \right)
\label{eqn:reduced_growth}
\end{equation}
where $f_{(c+b)}$ is the growth rate in only the CDM and baryon components, but \emph{calculated in the cosmology including the exotics}. We will call this the \emph{reduced growth rate}. The true growth rate is a small perturbation about this, with all potentially oscillatory contributions isolated. The effects of the axions and neutrinos contribute to this reduced growth rate only through the gravitational couplings via the potential $h$, and through the background expansion coming from $\mathcal{H}$.

It is this reduced growth rate that was shown in Fig.~\ref{fig:growth1}. In the blow up insert, we see that as expected it traces somewhat the average of the oscillations in the true growth rate, and shows scale dependence.

\begin{figure}
$\begin{array}{@{\hspace{-0.2in}}l}
\includegraphics[scale=0.4,trim=18mm 5mm 10mm 0mm,clip]{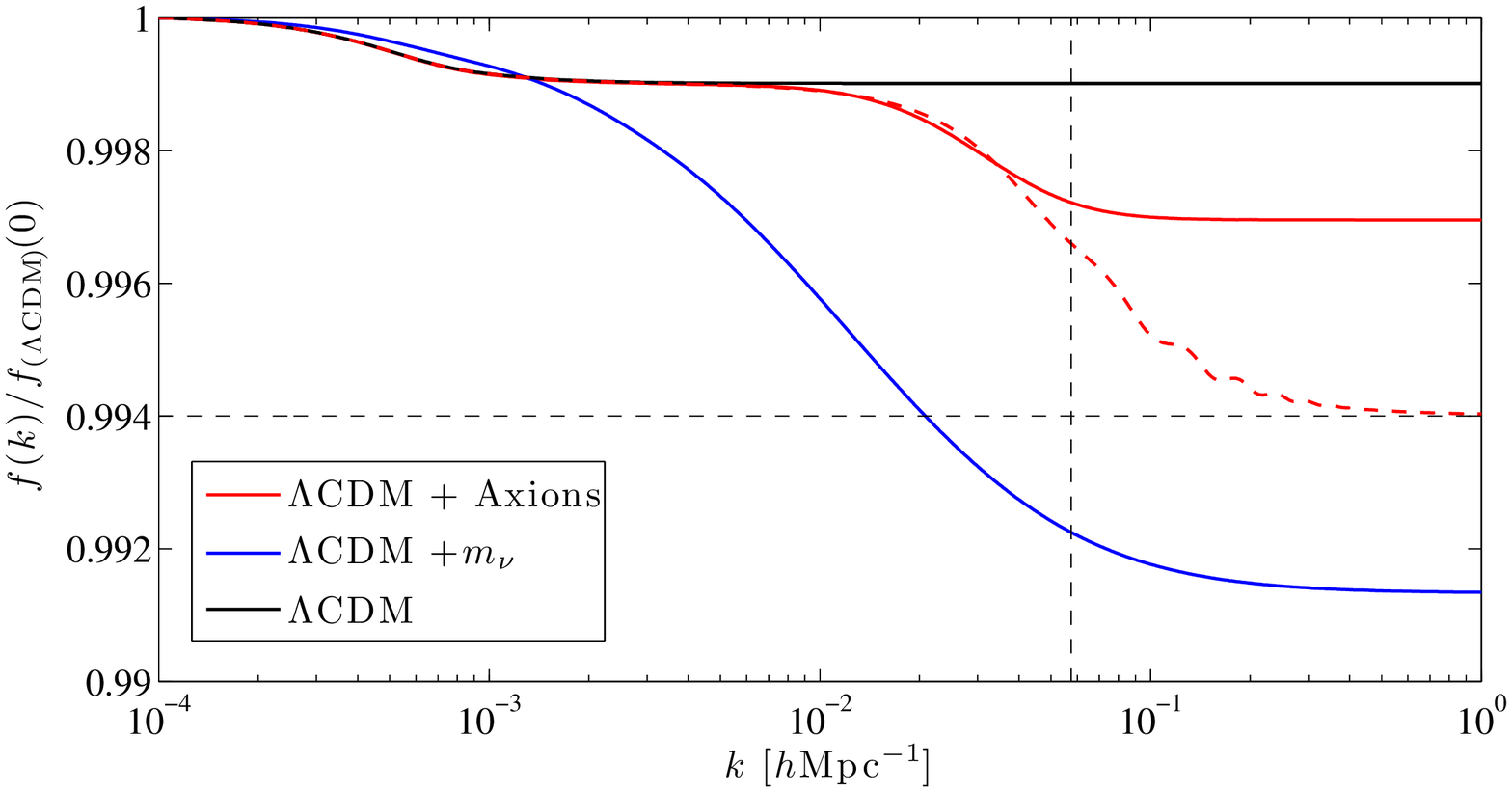}
\end{array}$
\caption{The reduced growth rate, $f_{(c+b)}(k)$, for three fiducial cosmologies: $\Lambda$CDM; $\Lambda$CDM with axions: $m_a=10^{-29}\unit{eV}$, $f_{ax}=0.01$; $\Lambda$CDM with massive neutrinos, $m_\nu=0.055\unit{eV}$, $N_{\mathrm{eff,mass}}=3.04$. We normalise all growth rates to one on the largest scales to account for normalisation by $\Lambda$CDM on the largest scales at arbitrary redshift. The results shown in solid are numerical, and the dashed line is that predicted by our fit for $\Delta f$ in the presence of axions, Eq. \ref{eqn:growthfit}. The vertical dashed line on the plot shows the expected value of $\bar{k}_m$, fit using the formulae of \cite{marsh2010}, while the horizontal dashed line shows the expectation of Eq. \ref{eqn:deltaf_app}.}
\label{fig:growth2}
\end{figure}

To more clearly show the scale dependence, in Fig.~\ref{fig:growth2} we plot the reduced growth rate, $f_{(c+b)}$ as a function of $k$, normalised to one on the largest scales so that $z$-dependence and normalisation to $\Lambda$CDM are absent . There is a clear step occurring at $k\approx \bar{k}_m$, the same scale as the suppression of power in $P(k)$. 

We have checked that the same growth rate is recovered in both the case where massive neutrinos are included exactly\footnote{In this case the growth rate contains no visible-by-eye oscillations, and is computed exactly using a numerical derivative of $\delta$.}, and where an averaging is taken over the oscillations present with an axion component. The accuracy of the approximation is to within a few percent of the total step size, which is itself only a few percent of the total value of the growth rate in a pure $\Lambda$CDM cosmology. It is thus this reduced growth rate that we will use in our forecasts for galaxy redshift surveys (GRS), which require the growth as a function of $k$ in a redshift bin, and is output for them from our modified version of CAMB. In these forecasts we therefore make no use of the fitting $f=\Omega_m(z)^{\gamma_g}$. 

Now we turn our attention to understanding the size, shape and position of this step in the growth rate in more detail. We have already described the effect of axions on the matter power spectrum using the transfer function $\tilde{T}_{ax}$. We use the same transfer function for the overdensities, i.e. $\delta_m=\tilde{T}_{ax}(k)\delta_{\Lambda \mathrm{CDM}}$. Note that this is smooth as a function of $z$: $T_{ax}$ is already an on-average fitting. Note also that this is an alternative to parameterisations of scale dependent growth used in discussing massive neutrinos in \cite{eisenstein1999,joudaki2011}. Substitution into Eqn.~\ref{eqn:growthdef} immediately yields:
\begin{align}
f&= f_{(\Lambda \mathrm{CDM})} + \frac{1}{\mathcal{H}}\frac{\dot{\tilde{T}}_{ax}}{\tilde{T}_{ax}} \nonumber \\
&= f_{(\Lambda \mathrm{CDM})} - \Delta f(k,z) \nonumber \\
\label{eqn:growthfit}
\end{align}

Using our fitting formulae we can calculate $\Delta f$. We work in the regime where $\tilde{f}_{ax}$ is fixed as a function of time, i.e. the axions have completed their transition to matter like behaviour, which will always be true for our fiducial models in the redshifts of interest, and in any case the fitting formulae break down where this condition is not satisfied. 

In the redshifts of interest, the $z$ dependence of the additional term is mild relative to $f_{(\Lambda \mathrm{CDM})}$, and so the shape of the growth rate as a function of $z$, modulo the oscillations, is largely unaltered by the presence of axions. In Fig.~\ref{fig:growth1} we have already shown the reduced growth rate as a function of $z$, and see that this is indeed the case. In Fig.~\ref{fig:growth2} we also show the fit of Eqn.~\ref{eqn:growthfit}. Notably, since the fit gives us the total growth, $f$, rather than the modified growth, the fit overestimates the size of the step. This can be taken as an indication of the amount by which the modified growth is perhaps an underestimate of the true step.

We estimate the change in growth rate amplitude at scales below $\bar{k}_m$ to be:
\begin{equation}
\Delta f(k>\bar{k}_m,z=0) = (1-q) \approx \frac{3}{5}f_{ax}
\label{eqn:deltaf_app}
\end{equation}
for small $f_{ax}$.

The most important feature of this analysis has been the identification of smooth, step-like, scale dependent growth occurring in models containing an ultra-light scalar field caused by the same physics as causes the suppression of power in the matter power spectrum, and also mimicking the corresponding effect due to massive neutrinos, though to a lesser degree, as seen in Fig.~\ref{fig:growth2}. While the size of this effect is not accurately estimated analytically, it's location in $k$-space can be predicted with reasonable accuracy. The location of the step depends only on the mass of axion, while the size depends only on the density fraction. Measurements of the growth rate amplitude will therefore measure the fraction in axions, but be insensitive to the mass unless scale dependent growth can be resolved at $\bar{k}_m$. Suppression of the growth rate, if measured in a non-scale dependent way, is clearly degenerate with the total total matter content, evident from the simple fitting in $\Lambda$CDM of $f=\Omega_m(z)^{\gamma_g}$, and suggesting further positive correlation between $f_{ax}$ and $\Omega_c h^2$. Again, using a single observable, precise scale dependent measurements are required to isolate a unique signal.

Weak lensing tomography also measures the growth rate, and we can use this to put tight constraints  on the existence of smooth components, such as axions at wavenumbers larger than $k_m$, or massive neutrinos at wavenumbers larger than $k_{FS}$, from the amplitude change relative to $\Lambda$CDM \cite{hu2002b}. In the case of weak lensing, which measures the growth rate more accurately and via different means, no approximation is made at all and the full numerical evolution of the overdensity is used. In this case, the oscillations are naturally smoothed by the redshift integral (see below).

\subsection{Galaxy Weak Lensing}
\label{weaklensing_theory}

Galaxy weak lensing was first observed in 2000 by \cite{waerbeke2000,kaiser2000,bacon2000,wittman2000}. The measurement of the weak lensing power spectrum is a direct probe of the dark matter. Through tomography, made possible by measuring galaxy photometric redshifts, we also gain information in the radial or temporal direction, and probe the growth rate and distance-redshift relation. This measurement of the growth rate gives constraining power for the presence of ultra-light axions. 

We first review some weak lensing basics. The effect of weak gravitational lensing is usually split into two components: the complex shear $(\gamma_1,\gamma_2)$ and the convergence $\kappa$, with the shear being derivable from the convergence (see \cite{Bartelmann2001} for a complete review). Let us denote $\chi$ the comoving distance and $r(\chi)$ the coordinate distance, defined by $r(\chi) = K^{-1/2}\sin\left(K^{1/2}\chi\right)$ for a closed universe, $r(\chi) = \chi$ for a flat universe, and $r(\chi) = (-K)^{1/2} \sinh\left((-K)^{1/2}\chi\right)$ for an open universe, where $K$ is the curvature.

The convergence in a given direction $\hat{\boldsymbol{n}}$ of the sky is given by an integral along the line-of-sight \cite{Huterer2002}
\begin{equation}
\kappa(\hat{\boldsymbol{n}},\chi) = \int_0^{\chi} W(\chi')\delta(\chi')\mathrm{d}\chi'
\label{eqn:convergence}
\end{equation}
where $\delta$ is the density perturbation and
\begin{equation}
W(\chi) = \frac{3}{2}\Omega_m H_0^2 g(\chi)(1+z).
\label{eqn:weightfunction}
\end{equation}
is a weighting function. $g(\chi)$ is given by
\begin{equation}
g(\chi) = r(\chi) \int_\chi^{\infty} \mathrm{d}\chi' n(\chi')\frac{r(\chi'-\chi)}{r(\chi')}
\label{eqn:gofchi}
\end{equation}
where $n(\chi)$ is the source distribution:
\begin{equation}
n(z) = \frac{3}{2z_0} \left(\frac{z}{z_0}\right)^2 e^{-(z/z_0)^{3/2}}
\label{eqn:nofz}
\end{equation}

We can now expand the convergence in multipoles $\kappa_{lm}$, and define the convergence power spectrum by the relation
\begin{equation}
\left\langle \kappa_{lm} \kappa_{l'm'} \right\rangle = \delta_{ll'}\delta_{mm'} P_l^\kappa,
\label{eqn:convergence_powspec_def}
\end{equation}
and using equation (\ref{eqn:convergence}) under Limber's approximation, we have the following expression:
\begin{align}
	P_l^\kappa ={}&  \int_0^{\chi_\infty}\mathrm{d}\chi \frac{W^2(\chi)}{r^2(\chi)} P(l/r(\chi),z) \\
	 ={}& \frac{9}{4}\Omega_m^2 H_0^4 \int_0^{\chi_\infty}d\chi \left(\frac{g(\chi)}{r(\chi)}\right)^2 (1+z)^2 P(l/r(\chi),z)\nonumber
\label{eqn:convergence_powspec}
\end{align}
where $\chi_\infty$ stands for $\chi(z\rightarrow\infty)$, and $P$ is the matter power spectrum.

In this way, the convergence power spectrum depends directly on the matter power spectrum at redshift $z$, and thus tomographic information will measure the growth rate. The scale at which the growth rate is measured depends on the pixel size, and for our purposes this will measure \emph{below} either $k_m$ or $k_{FS}$ for \emph{all} axion and neutrino masses considered. Therefore, the constraint on these species will be independent of this scale and will depend only on the density fraction, i.e. for the case of ultra-light axions where the density fraction depends only on the initial misalignment angle \emph{constraints on $f_{ax}$ from weak lensing are expected to be independent of axion mass}, but due to their similar effects will be degenerate with the neutrino density fraction, and the dark energy equation of state, the strongest constraint being from the amplitude of the growth rate \cite{hu2002b}. In addition, our method for calculating the weak lensing tomography comes directly from integrating the density parameters inside CAMB, and so is not limited by the approximation of the reduced growth rate used for GRS, as mentioned already above. 

Note that this effect in the growth rate is distinguishable from many DE effects, which occur predominantly through modifying the distance-redshift relation via changes in the expansion rate:
\begin{equation}
\chi (z) = \int_0^z \frac{dz'}{H(z')}
\end{equation}
As we have already seen, axion and neutrino effects on the expansion rate during the matter era are small to non-existent. Establishing the distance-redshift relation in addition to the growth rate using lensing tomography serves to break degeneracies occurring between axion/neutrino components and DE. For example, as discussed earlier one can change the DE contribution to restore $k_{eq}$ and the canonical shape of $P(k>k_{eq})$, but this then demands a change in $w$ if we are also to restore the Hubble expansion, leading to an apparent degeneracy between $w$ and $m_\nu$ or $f_{ax}$, which can be broken by lensing tomography or BAO measurements that pin down the expansion as a function of redshift \cite{hannestad2006}. 

Because of the dependence on $P(k,z)$, all of our intuition of the similarity of axion and neutrino effects on the matter power spectrum should carry over into galaxy weak lensing. Only our Fisher matrix analysis will show what degeneracies really exist in the fiducial models we investigate under the precision of the observations in question.

\begin{figure}
\includegraphics[scale=0.85,trim=0mm 5mm 10mm 0mm,clip]{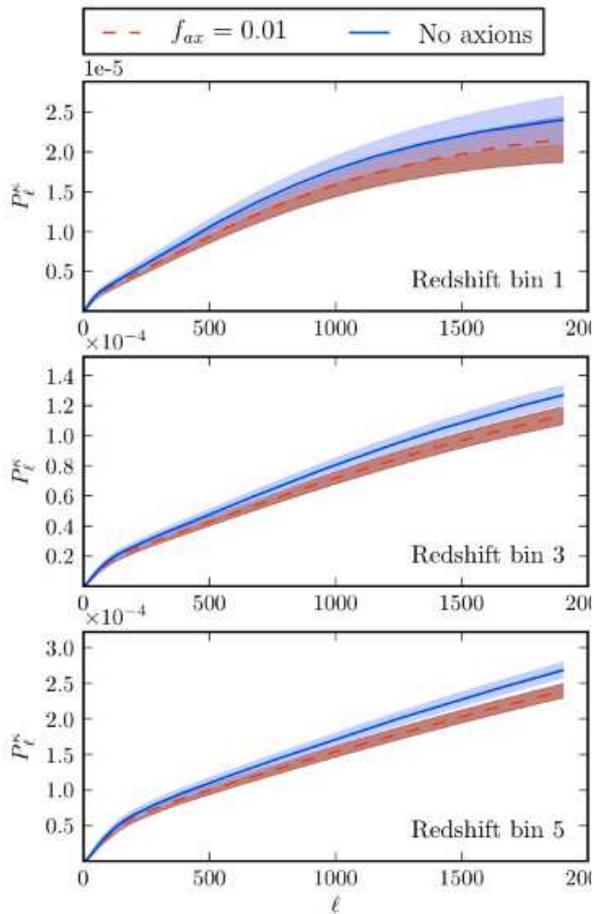}
\caption{The convergence power spectrum for three redshift bins, with expected $1\sigma$ errors from a large future weak lensing survey, defined in Appendix~\ref{appendix_survey_details}. Overlapping regions of error are shown in purple. The models compared are $\Lambda$CDM, and $\Lambda$CDM$+f_{ax}$, with $m_a=10^{-29}\unit{eV}$, $f_{ax}=0.01$.}
\label{fig:galaxy_lensing1}
\end{figure}

In Fig. \ref{fig:galaxy_lensing1} we show the effect in the convergence power spectrum of adding an axion fraction $f_{ax}=0.01$ in a species of mass $m_a = 10^{-29}\unit{eV}$ in various different redshift bins. The unclustered species causes a suppression of power on small scales, which increases with increasing fraction in that species, as one would expect from the structure suppression in $P(k)$. The errors on the convergence power spectrum expected form a large future survey are also shown (see Appendix~\ref{appendix_survey_details} for survey parameters). In some single redshift bins, these are not strong enough to distinguish a $\Lambda$CDM model from a model with axions at $1\sigma$, while in others they just are. The amplitude of suppression of power is independent of axion mass, and like effects in the growth rate is smaller than the same effect due to massive neutrinos.

A possible uncertainty on constraints from weak lensing lies in the treatment of the non-linear regime, necessary for weak lensing calculations. Inside CAMB we use a standard version of halofit \cite{halofit}, which is optimised from simulations to standard CDM and may not accurately describe the non-linear effects of axions or massive neutrinos. However, mistreatment of non-linear effects likely leads to an underestimate of the suppression of power on small scales \cite{joudaki2011}, and so our projected weak lensing constraints should be conservative. Hannestad et al \cite{hannestad2006} estimate that the effect of assuming  no non-linear effects in neutrinos leads to an uncertainty in lensing observables of around 0.1\%. For a discussion of the non-linear effects of neutrinos see \cite{saito2008}, and for an N-body simulation with ultra-light scalars, see \cite{woo2009}.  

\subsection{The CMB}
\label{cmb}

\subsubsection{Temperature Power Spectrum}

Here we discuss axion effects on the CMB temperature-temperature auto correlation power spectrum, $C_\ell^{TT}$, the TT power spectrum (for a review of CMB anisotropies, see \cite{hu2002}, and for specific applications to HDM and neutrinos see \cite{dodelson1995,bashinsky2004}). 

\begin{figure*}
\centering
$\begin{array}{@{\hspace{-0.2in}}l@{\hspace{-0.5in}}l}
\includegraphics[scale=0.35,trim=20mm 0mm 10mm 0mm,clip]{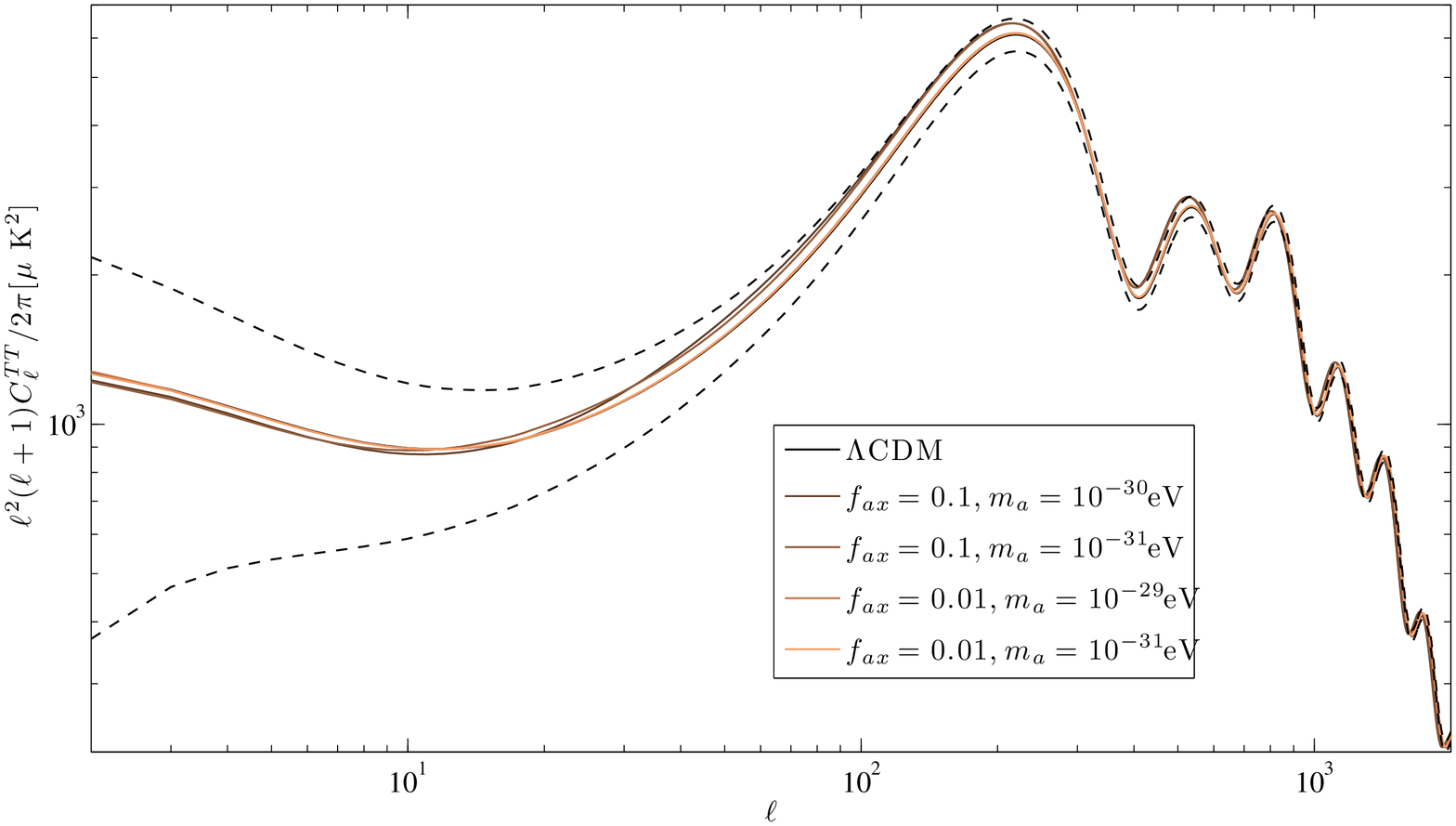}&
\includegraphics[scale=0.35,trim=20mm 0mm 10mm 0mm,clip]{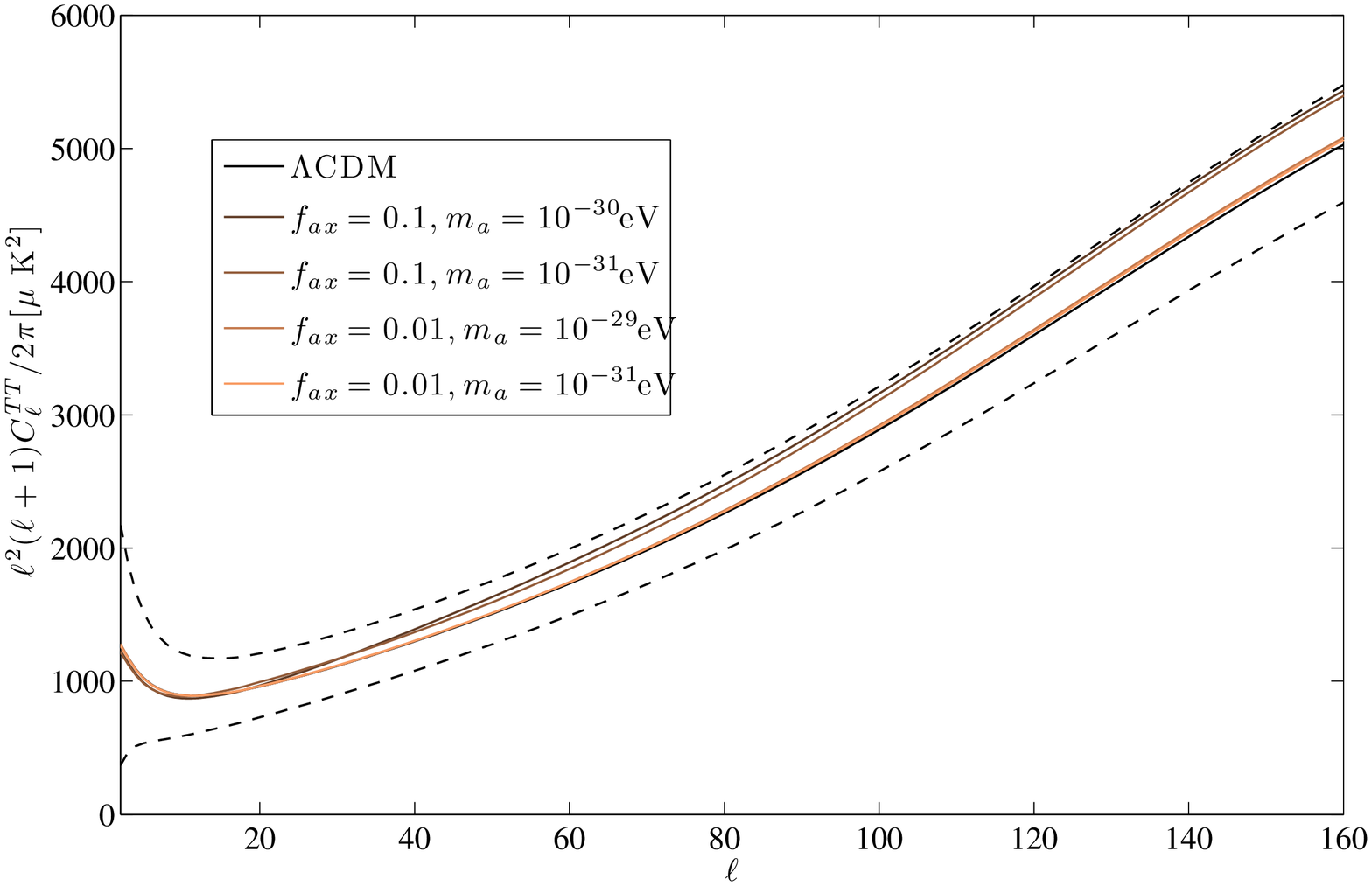}\\[0.0in] 
\end{array}$
\caption{Effect of axions on the CMB power spectrum. Left panel: TT power spectrum. Right Panel: magnified low $\ell$ region showing the ISW effect. Expected error bands from Planck are shown as dotted lines, see Section~\ref{sec:cmb_forecasts}.}
\label{fig:cell_ax}
\end{figure*}

In Fig. \ref{fig:cell_ax} we show the effects of various axion fractions and masses on $C_\ell^{TT}$. Axions have the effect of shifting the acoustic peaks by changing the epoch of equality. The changes in equality in these models, $\Delta z_{eq}(f_{ax}) = z_{eq}(0)-z_{eq}(f_{ax})$, are $\Delta z_{eq}(0.1)=328$ and $\Delta z_{eq}(0.01)=32.2$ respectively. 

The most pronounced effect for heavier axions occurs in the Integrated Sachs-Wolfe (ISW) region at very low $\ell$, as pointed out in \cite{amendola2005}. As also pointed out in \cite{amendola2005}, this ISW effect is maximal for masses at the larger end of those we study, suggesting potentially strong CMB constraints caused by this. These heavier axions, as discussed earlier, are making their transition from cosmological constant to CDM behaviour very close to CMB formation. As such, they are contributing a significant fraction to the energy density while in a non-standard stage of evolution, and thus causing large variations in the potential on the largest scales where this transition is beginning. At high $\ell$ there is no difference in the effect of axions of different mass, so small scale CMB measurements will constrain the total fraction possible in ultra-light axions independent of mass. Heavier axions than those we consider here that have completed their transition to CDM in the radiation dominated era should be expected to have no effect at all on the CMB, as they are gravitationally indistinguishable from CDM.

In addition, the low-$\ell$ ISW effect distinguishes axions from massive neutrinos. It is in the high-$\ell$ region of the TT power spectrum that evidence from the CMB for extra relativistic energy density comes, due to the effect of this energy density on the pre-recombination expansion rate and corresponding increase in the amount of Silk damping leading to the required lower fluctuation amplitude \cite{dunkley2010,hou2011}. The best way to explore the effect of extra relativistic species on the TT power spectrum is to hold fixed a number of quantities and isolate the extra effects at high-$\ell$. We can apply the same logic to further understand the effect of ultra-light axions since they affect the epoch of equality in the same way as extra relativistic energy density: increasing $N_{\mathrm{eff,rel}}$ means more radiation, making equality later, while introducing axions with $z_{\mathrm{osc}}<z_{\mathrm{eq}}$ reduces the amount of dark matter at equality, giving the same effect \cite{bashinsky2004}. The amount of baryons, $\rho_b$, should be fixed to keep the even and odd acoustic peaks in the correct ratio of amplitude. The epoch of equality, $z_{\mathrm{eq}}$, being tightly constrained by the position of the first peak should then be fixed by changing $\rho_d$. The angular size of the sound horizon, $\theta_s = r_s/D_A$, should be fixed by changing $H_0$. Finally, the height of the first peak, which isolates the low and high-$\ell$ effects, should be fixed by changing the scalar amplitude, $A_s$. 

When all of these changes are made, the result is to completely remove the effect of axions from the CMB, while the same changes made in a cosmology $\Delta N_{eff}>0$ leave the expected effect in the Silk damping tail This is because the extra relativistic energy density has left an imprint in altering the expansion rate during the radiation era. In an axion cosmology, restoring equality, as mentioned above in our discussion of the BAO, also restores the Silk damping scale. Axions in the CMB are degenerate with a combination of changing $\rho_d$ and $H_0$, while neutrinos are not. These observations are important, since they will apply also to massive neutrinos: \emph{the CMB breaks the degeneracy between axions and neutrinos due to neutrino effects on the expansion rate in the radiation era, and axion ISW effects in the matter era}.

E-mode CMB polarisation auto-correlation power spectra , $C_\ell^{EE}$, and temperature-polarisation cross-correlations, $C_\ell^{TE}$, the EE and TE power spectra, are also  useful cosmological probes, and they too are used in our forecasts. Polarisation will also be particularly important if ultra-light axions have the model dependent coupling to $\vec{E}\cdot\vec{B}$. In this case the polarisation angle can be rotated by $\Delta \beta \approx 10^{-3}$, an effect within reach of Planck and CMBPol. If an observation of this rotation were combined with other axion observations as described above, this would provide very strong evidence for the existence of axions with this coupling, but since the coupling is model dependent, polarisation rotation need not accompany other axion effects. Observation of polarisation effects without other axion effects would, however, rule out axions as a cause of the rotation. We do not include the possibility for such a rotation in our models.

The CMB is an important constraint on another effect of axions, which we do not consider in this work: axion isocurvature perturbations. Axion isocurvature contributions are significant if the energy scale of inflation is high. The authors of \cite{axiverse2009} assert that a low scale of inflation is natural in the axiverse scenario. If axions are not to overproduce isocurvature perturbations, the energy scale of inflation must be so low that gravity waves would not be significantly produced. The authors of \cite{fox2004} propose this as a direct method of falsifying the existence of string axions, while the authors of \cite{mack2009a,mack2009b} propose using inflationary parameters in conjunction with axion isocurvature contributions as a method of quantifying fine tuning within string axion models. We propose to investigate axion isocurvature in detail in a forthcoming paper.




\subsubsection{CMB Lensing}

CMB lensing has recently been detected and measured by ACT \cite{das2011,sherwin2011}, and its observation by Planck will be a powerful cosmological probe.

CMB lensing, like galaxy weak lensing, is a direct probe of the DM density and expansion history between the surface of last scattering and us. Hence, in terms of CMB observables, the EDE fraction and $\Sigma m_\nu$ are most strongly constrained by CMB lensing \cite{joudaki2011}. However for our purposes, we cannot include CMB lensing and galaxy weak lensing together since addressing correlations between these observables is beyond the scope of this paper. We do not use CMB lensing in our forecasts, but we show the power spectrum here and make some comments for the sake of completeness.

\begin{figure}
\centering
$\begin{array}{@{\hspace{-0.2in}}l}
\includegraphics[scale=0.45,trim=27mm 0mm 10mm 0mm,clip]{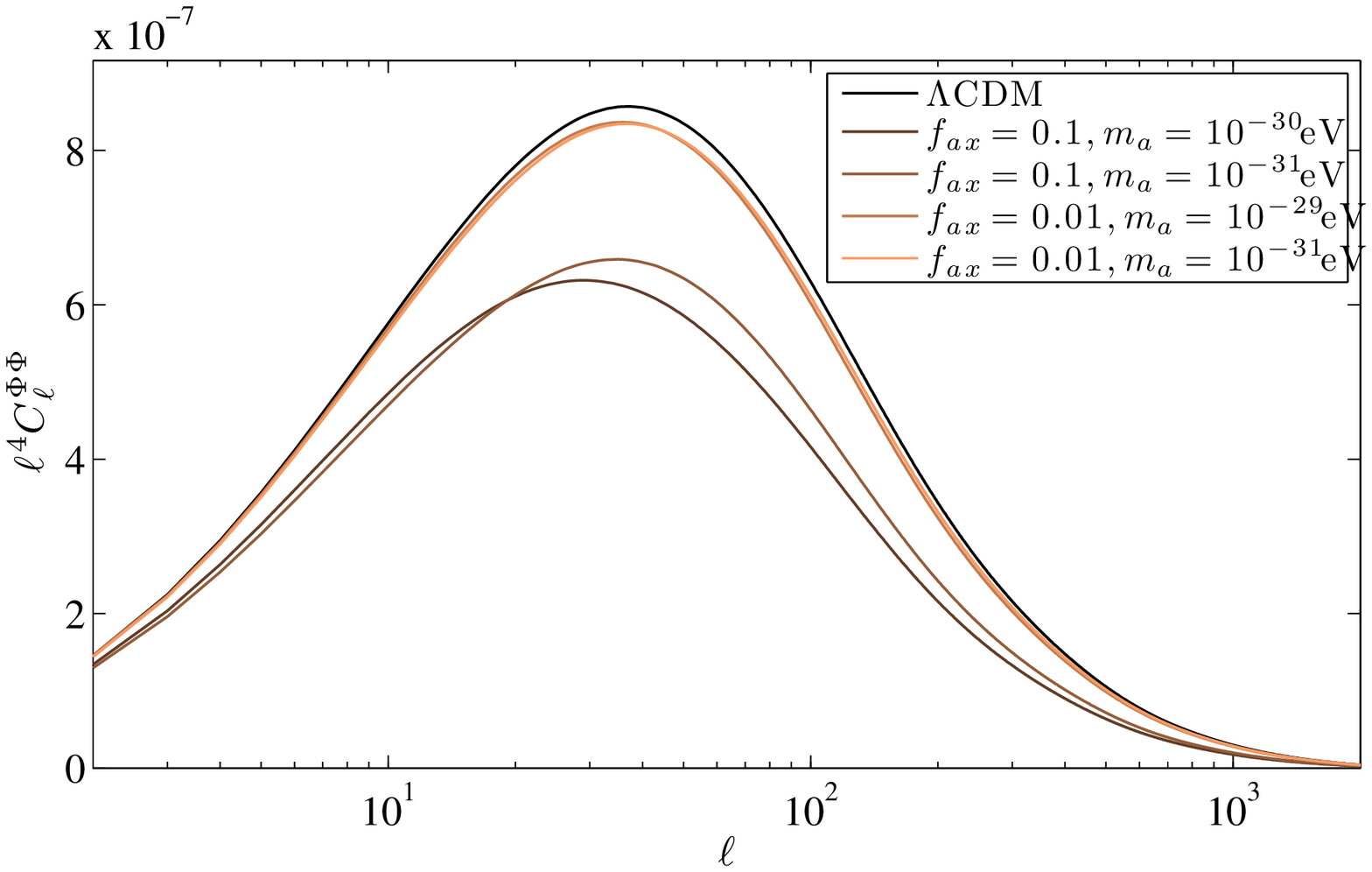}
\end{array}$
\caption{The CMB lensing power spectrum.}
\label{fig:cpp_ax}
\end{figure}

We show the lensing power spectrum in Fig.~\ref{fig:cpp_ax}. A large fraction in axions produces a noticeable effect in reducing the lensing power. However, if as above we fix $z_{eq}$ and $\theta_S$ again the effect can be totally removed. This occurs due to the degeneracy of axions with other parameters in their effects on the matter power spectrum, as already discussed.

%
%
\section{Cosmological Observables: Forecasts}

In this section we consider the potential to identify traces of axion dark matter with some of the cosmological probes of the upcoming decade.   We adopt a Fisher matrix approach to forecast results achievable with a next generation galaxy redshift and weak lensing survey.  There are many large weak lensing and galaxy redshift surveys proposed for the coming years; we choose to model our surveys to be similar to the proposed Euclid mission \cite{euclidRB,euclidWEB}.  We combine these results with Fisher forecasts for an all sky CMB survey based on the Planck mission.   The Fisher matrix provides the lowest possible intrinsic statistical uncertainty, thus results from Fisher forecasts tend to be optimistic when compared to real results.  We describe in detail our own implementation in galaxy redshift surveys and galaxy weak lensing tomography, while our CMB forecasts are made using the pre-existing package ``FisherCodes''.    

Fisher matrices are widely used in cosmology to forecast uncertainties in parameters $\Theta$ of an underlying model of an observable $x$ given an uncertainty in a measurement of $x$, $\sigma_x$. If we assume that the uncertainties in our model are Gaussian, the Fisher matrix can be calculated directly from derivatives of the model, and is

\begin{equation}
F_{ij}=\frac{1}{\sigma_x^2}\frac{\partial x}{\partial \Theta_i} \frac{\partial x}{\partial \Theta_j} 
\label{direct_calc_FM}
\end{equation}

The Fisher matrix (FM) is the inverse of the covariance matrix (CM) of the parameters.  Working with the FM is particularly useful since it allows forecasted constraints from different surveys $A$ and $B$ to be easily combined to provide combined forecasted uncertainties from both surveys, so that $F_{ij}^{Total}=F_{ij}^{A}+F_{ij}^{B}$. In the cases we consider here, the observable is generally a power spectrum (multipole spectrum for the CMB, galaxy power spectrum for the galaxy redshift survey, and convergence power spectrum for weak lensing).  
 
 \subsection{Forecasting For Galaxy Redshift Surveys}

Tegmark \cite{1997PhRvL..79.3806T} showed that for a galaxy redshift survey, the Fisher matrix is given by:
\begin{equation}
F_{ij}=\int_{\vec{k}_{\text{min}}}^{\vec{k}_{\text{max}}} \frac{d^{3}\vec{k}}{2 (2\pi)^3} \left( \frac{\partial \ln P(\vec{k})}{\partial \Theta_i}  \right) \left( \frac{\partial \ln P(\vec{k})}{\partial \Theta_j}  \right) V_{\text{eff}}(\vec{k})
\label{galaxy_FM_vector_k}
\end{equation}

The `effective volume', $V_{\text{eff}}$, essentially provides a weighting for the uncertainty in each $P(k)$, and (assuming constant galaxy density) is given by:
\begin{equation}
V_{\text{eff}}(k) = V_{o}  \left( \frac{nP(k)}{1+nP(k)}  \right) ^2
\label{V_eff}
\end{equation}
where $V_o$ is the volume of the survey and $n$ is the galaxy density. The power spectrum that is measured is that of the galaxies, not the underlying dark matter. The two can be related by including the effects of bias and redshift space distortions. The galaxy bias $b$ is defined as $\delta_g$=$b\delta_m$, where $\delta_m$ is the matter density fluctuation, and $\delta_g$ is galaxy density fluctuation. Furthermore, the true radial position of the galaxies as measured by the redshift is distorted by the peculiar velocities of the galaxies.  On large scales, coherent infall into overdensities causes an apparent compression along the line of sight, which depends on the growth rate, $f_g$ \cite{1987MNRAS.227....1K}.  On small scales, large peculiar velocities due to thermal motions in virialised clusters can elongate the positions of galaxies along the line of sight. The biased, redshift space galaxy power spectrum we used is based on the model used by \cite{2010PhRvD..81d3512S}, and is given by
:
\begin{equation}
P(k,\mu) =b^2  \left( \frac{1}{1+(k \mu \sigma_v)^{2} /2} \right)  \left( 1 + \frac{f_g}{b}  \mu^2 \right)^2  P(k)
\label{model_pk}
\end{equation}
 where $\mu$ is the cosine of $k$ modes parallel and perpendicular to the line of sight. The $b^2$ term accounts for the linear galaxy bias, and the $\sigma_v$ term introduces small scale velocity dispersion. The $f$ term models the large scale growth. The Fisher matrix can now be written in terms of $\mu$ and the magnitude of $\vec{k}$.  

\begin{align}
F_{ij}=\int_{-1}^{1}\int_{k_{\text{min}}}^{k_{\text{max}}} \frac{2\pi dk d\mu}{2 (2\pi)^3} &\left( \frac{\partial \ln P(k,\mu)}{\partial \Theta_i}  \right) \nonumber\\
 &\left( \frac{\partial \ln P(k,\mu)}{\partial \Theta_j}  \right) V_{\text{eff}}(k)
\label{galaxy_FM_vector_k}
\end{align}

We choose  a model to be similar to the Euclid galaxy redshift survey, consisting of 18,000 square degrees between redshift 0.5 and 2.1, in 15 independent redshift bins \cite{Euclid_science_requirements_document_2011}. Due to uncertainty at large scales, the Fisher matrix is insensitive to the choice of $k_{\text{min}}$.  Following \cite{2010MNRAS.409..737W}, we set $k_{\text{max}}$ so that the variance is given by $\sigma^2( \frac{\pi}{2k_{\text{max}}} )=0.25$, and also that $k_{\text{max}}<0.2h^{-1}$Mpc at higher redshifts.  We assume a constant galaxy density of $1\times10^{-3}  h^{-3}$Mpc$^3$ and a constant linear bias of 1.7. The volume between each bin for a survey size of 18,000 square degrees was calculated according to \cite{1999astro.ph..5116H}, with code implemented by \cite{2009arXiv0906.0974B}. Full details of the redshift bins is included in  Appendix~\ref{appendix_survey_details}. To calculate the total Fisher matrix for each survey, individual Fisher matrices were calculated for each redshift bin, and then summed.

For each fiducial model, a Fisher matrix was calculated for each of the 15 independent redshift bins in the survey model. These matrices were summed to obtain the Fisher matrix for the whole redshift survey.

\subsection{Forecasting For Galaxy Weak Lensing}

As shown in \cite{Hu1999}, the Fisher matrix for the convergence power spectrum $P_l^\kappa$ is given by:
\begin{equation}
F_{ij}^{WL} = \sum_{\ell_{min}}^{\ell_{max}} \frac{\ell + 1/2}{f_{sky} \left(P_l^\kappa + \langle \gamma_{int}^2\rangle/\bar{n}\right)^2} \frac{\partial P_l^\kappa}{\partial \Theta_i} \frac{\partial P_l^\kappa}{\partial \Theta_j}
\label{eqn:convergence_FM}
\end{equation}
where $f_{sky}$ is the fraction of the sky covered by a survey, $\bar{n}$ is the average number density of galaxies, and $\langle\gamma_{int}^2\rangle^{1/2}$ is the galaxy intrinsic rms shear. We took $\langle\gamma_{int}^2\rangle^{1/2} = 0.22$ (following \cite{bernardis2009}) to match the Euclid experiment, $f_{sky} = 0.5$ and $\bar{n} \simeq 4.18\times10^{8} \unit{\text{sr}^{-1}}$, and the galaxy distribution function given by Eqn.~\ref{eqn:nofz}, so that $\int_0^\infty n(z) \mathrm{d}z = 1$, with $z_0 = 0.9$.

However, since we are interested in Euclid-like surveys, we used weak lensing tomography which has been proven to improve significantly the constraints on cosmological parameters \cite{hu2002b}. We used 5 equal-galaxy-number bins, for $z = 0 - 3$. To account for the photometric redshift uncertainties, we followed \cite{Ma2005}. In the $i^\text{th}$ bin, limited by $z_i$ and $z_{i+1}$, we used the distribution $n_i(z)$ defined by
\begin{equation}
n_i(z) = \frac{1}{2}n(z) \left( \text{erf}\left( \frac{z_{i+1}-z}{\sqrt{2}\sigma_z} \right) - \text{erf}\left( \frac{z_{i}-z}{\sqrt{2}\sigma_z} \right) \right)
\label{eqn:niofz}
\end{equation}
with $\sigma_z = 0.03 (1+z)$.

In the convergence power spectrum, this will give a factor $W_i(\chi) W_j(\chi)$ instead of $W^2(\chi)$, where $W_i(\chi)$ comes from $g_i(\chi)$, itself defined with $n_i(z)$. Instead of a single power spectrum $P_l^\kappa$, we now have a matrix $\boldsymbol{\mathrm P}_l^\kappa$, where the $(i,j)$ component comes from the $W_i(\chi) W_j(\chi)$ factor.

Equation (\ref{eqn:convergence_FM}) can be generalized by
\begin{equation}
\label{eqn:tomography_FM}
F_{ij}^{WL} = \sum_{\ell_{min}}^{\ell_{max}} (\ell+1/2) f_{sky}) \text{Tr}\left( \boldsymbol{\mathrm C}_l^{-1} \frac{\partial \boldsymbol{\mathrm P}_l^\kappa}{\partial \Theta_i} \boldsymbol{\mathrm C}_l^{-1} \frac{\partial \boldsymbol{\mathrm P}_l^\kappa}{\partial \Theta_j} \right)
\end{equation}
where $\boldsymbol{\mathrm C}_l$ is given by $C_l^{ij} = P_l^{ij}+\langle \gamma_{int}^2 \rangle / \bar{n}_i \delta_{ij}$, and $\bar{n_i}$ is the average number of galaxies in the $i^\text{th}$ bin.

This expression does not take into account various other systematics. For instance, intrinsic alignments of galaxies can cause a false positive of the lensing effect: the ellipticities of nearby galaxies can be correlated because of gravitationnal interaction \cite{Heymans2003}. We did not investigate other sources of errors, such as the finite size effects of the point spread function or its anisotropy \cite{Huterer2006}.

The details of our weak lensing survey are given in Appendix~\ref{appendix_survey_details}.

\subsection{CMB Forecasts}
\label{sec:cmb_forecasts}

We use ``FisherCodes'' by Sudeep Das \cite{FisherCodes} to forecast results that could be obtained from the CMB.  We set the survey specifications for a Planck-style mission based on \cite{2006astro.ph..9591A}.  TT, TE and EE spectra were used, to $\ell_{max}=2000$, with $f_{sky}=0.8$.  Three channels at 100, 143, 217 GHz were used, with beam full-width half-maximum set to 9.5', 7.1' and 5.0'.  The temperature noise per pixel was set to 2.5, 2.2 and 4.8, and the polarization noise per pixel was set to 4, 4.2 and 9.8 $\times 10^{6}\Delta_T / T_{CMB}$.

\subsection{Results: CMB, Galaxy Redshift Survey and Weak Lensing Forecasts}
\label{results}

\begin{figure}
\centering
\includegraphics[width=9cm]{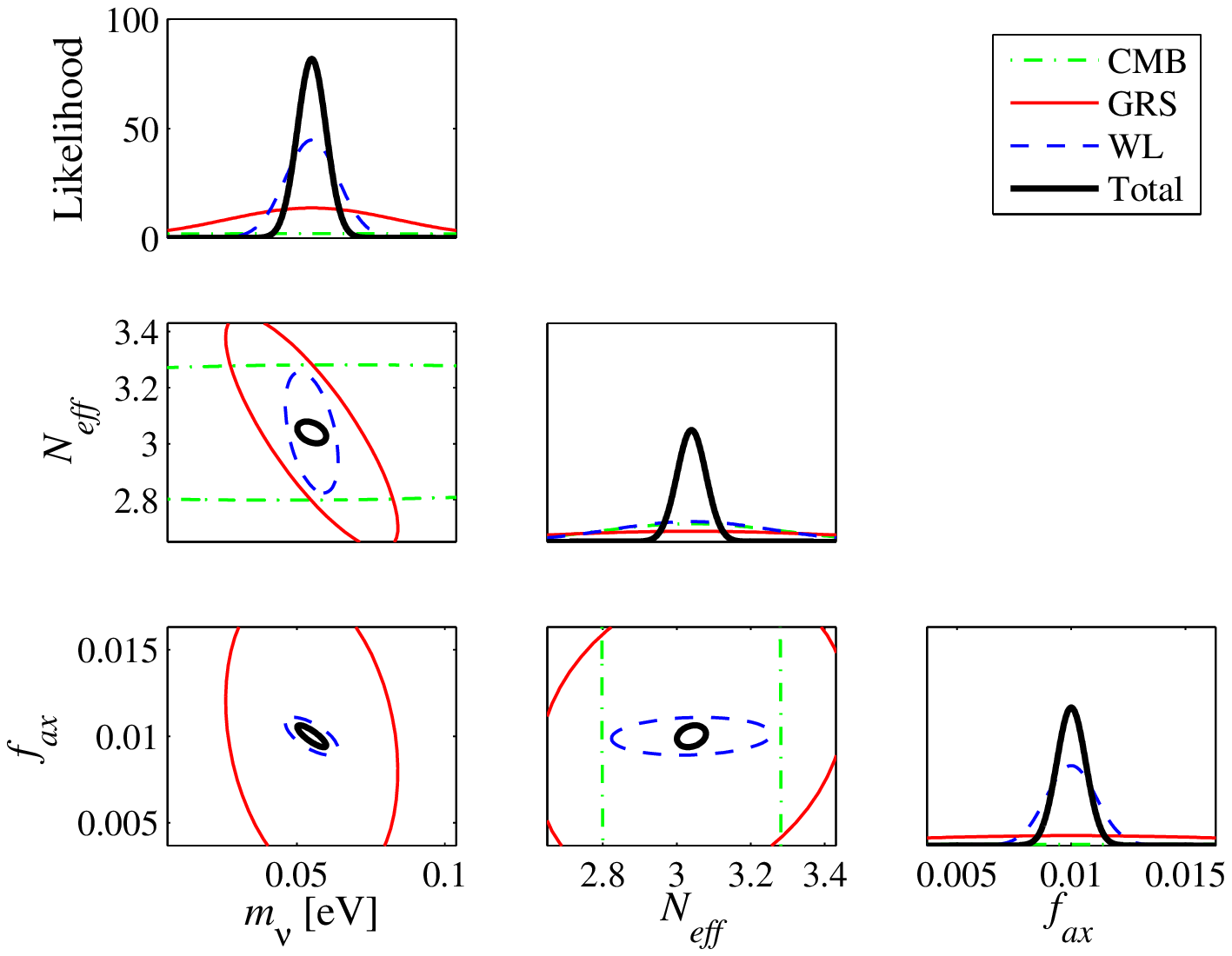}
\caption{One standard deviation Fisher ellipses for  $m_{\nu}$, $N_{eff}$ and $f_{ax}$.  The full combined set of parameters also includes  $w$, $\Omega_{b}h^{2}$, $\Omega_{c}h^{2}$, $n_s$ and $A_s$, which have been marginalized over in this plot.  Here the weak lensing forecast (blue dashed line) provides the best constraints. These forecasts are for an axion mass of $m_a=10^{-30}\unit{eV}$.}
\label{CornerTrianglePlot} 
\end{figure}

We first consider a cosmology with massive neutrinos and a fraction of CDM in axions.  We assume a fiducial model with parameters $w=-1$, $\Omega_{b}h^{2}=0.02258$, $\Omega_{c}h^{2}=0.1109$, $n_s = 0.963$, $A_s = 2.3\times10^{-9}$, $m_{\nu}=0.055$eV, $N_{eff}=3.04$ and $f_{ax}=0.01$.  We use a fixed value of $H_0=71.9$km s$^{-1}$Mpc$^{-1}$. Following \cite{2006astro.ph..9591A}, we also include the optical depth to reinozation $\tau=0.166$ and helium fraction $Y_{He}  = 0.24$ as free parameters in the CMB FM. Similarly, galaxy bias $b=1.7$ and non-linear velocity dispersion $\sigma_v=350$ kms$^{-1}$ were included as free parameters in the GRS  FM.  These parameters were marginalized over before the matrices were combined.  CMB, GRS and WL Fisher matrices were calculated for this fiducial cosmology for $m_a=10^{-29},10^{-30},10^{-31}$ and $10^{-32}$ eV.  As an example, Fisher ellipses for the $m_a=10^{-30}$ eV case are plotted in Fig. \ref{CornerTrianglePlot} for axion and neutrino parameters. We see the expected negative correlation between  $f_{ax}$ and $m_\nu$, but a small positive correlation between $f_{ax}$ and $N_{eff}$ in GRS, since in the case of massive neutrinos $N_{eff}$ adds more matter at late times.

\begin{figure}
\centering
\includegraphics[width=9cm]{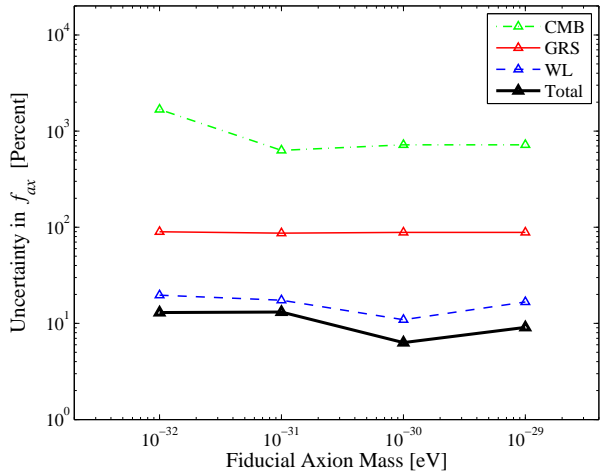}
\caption{Marginalized uncertainty in $f_{ax}$ for our three observables, evaluated for four different fiducial axion masses, for the cosmology $\Lambda$CDM$+f_{ax}+m_\nu$.  The uncertainty from the GRS and WL surveys does not change appreciably across the range of axion mass, whereas the uncertainty from the CMB survey decreases for higher axion masses. We cannot be sure if the small decrement in uncertainty in the WL survey at $m_a=10^{-30}$ is significant.}
\label{fAxUncert_vs_mAx} 
\end{figure}

With the different sets of forecasts over a range of axion mass, we can compare how the fully marginalised uncertainty on our parameters varies with axion mass.  We can see in Fig. \ref{fAxUncert_vs_mAx} that the uncertainty in $f_{ax}$ from  galaxy redshift surveys does not change appreciably with axion mass, as expected for the masses under consideration, and providing a good test that uncertainties in the numerical background evolution, which would vary with mass, do not affect the constraints. However, we do see some unexpected variation in the weak lensing constraints with mass, of $\mathcal{O}(1)$ of the total uncertainty. We believe this is due to numerical uncertainty in the complex calculation for lensing with an oscillating background, and so our results here can only be considered reliable to within this extra uncertainty. 

The CMB becomes more sensitive to $f_{ax}$ for higher mass: this we believe to be physical. It is caused by the increased ISW effect at redshifts close to the surface of last scattering, as expected. The trend to slightly better constraints at higher mass has been tested and seen to be stable: stability requires calculations to be made at extremely high accuracy. The required level of accuracy is computationally very expensive, and ensuring its stability has been the main limitation on models we have been able to forecast for. The constraining power of the CMB alone allows $f_{ax}\approx 0.1$ to be detected, which is consistent with the results of \cite{amendola2005}.

\begin{figure}
\centering
\includegraphics[width=9cm]{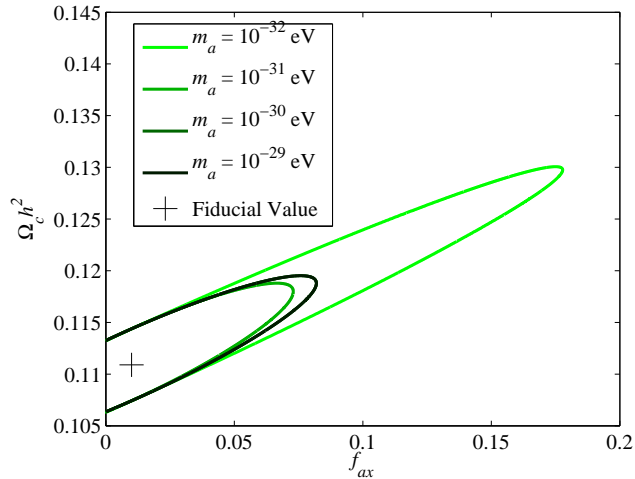}
\caption{Fisher ellipses for our forecast CMB survey for parameters $\Omega_{c}h^{2}$ and $f_{ax}$ for a range of axion mass. The ellipses for $m_a=10^{-29}\unit{eV}$ and $m_a=10^{-30}\unit{eV}$ lie directly on top of each other with the thickness of line in the plot. Note that if we fix $f_{ax}$, we obtain the same uncertainty on $\Omega_{c}h^{2}$ for all the different masses.}
\label{Oc_fAx_CMB_ellipses} 
\end{figure}

The improvement in the CMB measurement of $f_{ax}$ is shown again in Fig. \ref{Oc_fAx_CMB_ellipses} for the range of fiducial $m_a$, where we show the Fisher ellipse with $\Omega_c h^2$.These results show the expected positive correlation between $f_{ax}$ and $\Omega_c h^2$ caused by the effects on equality already discussed. The increase in CMB constraining power at high masses does not reach a sufficient level to break this degeneracy.
 
\begin{figure}
\centering
\includegraphics[width=9cm]{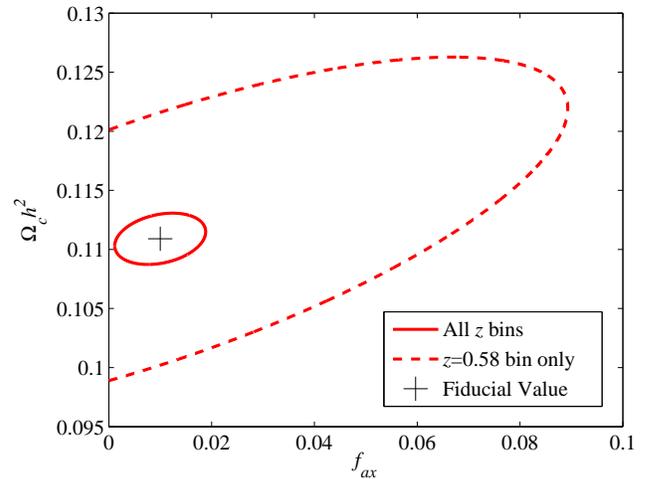}
\caption{Forecasts on $\Omega_{c}h^{2}$ and $f_{ax}$, for the galaxy redshift survey.  With one redshift bin of the survey (dashed line), the survey cannot discriminate between a cosmology with or without axions, and can only rule out $f_{ax}\lesssim10\%$.  When constraints from all 15 redshift bins are combined, detecting $f_{ax}$ \textgreater 0 is just possible to 1$\sigma$. The constraints are the same for all axion masses.}
\label{Oc_fAx_GRS_ellipses} 
\end{figure}

In Fig. \ref{Oc_fAx_GRS_ellipses} we show the effect of having redshift information on the constraints from a galaxy survey. Although a measurement of the power spectrum at a single redshift cannot distinguish $f_{ax}=0.01$ from $f_{ax}=0$ at $1\sigma$, this becomes possible when all $z$ bins are combined. We reiterate that this constraint is independent of axion mass.

\begin{figure}
\centering
\includegraphics[width=9cm]{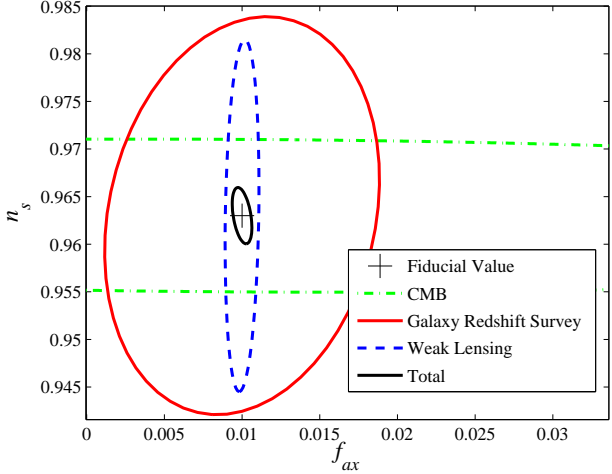}
\caption{Constraints on $n_{s}$ and $f_{ax}$, for  $m_a=10^{-30}\unit{eV}$, showing the expected positive correlation from galaxy redshift surveys. At this level of constraint, no correlation is visible in the CMB.}
\label{Ellipse_fAx_ns_AllObs} 
\end{figure}

\begin{figure}
\centering
\includegraphics[width=9cm]{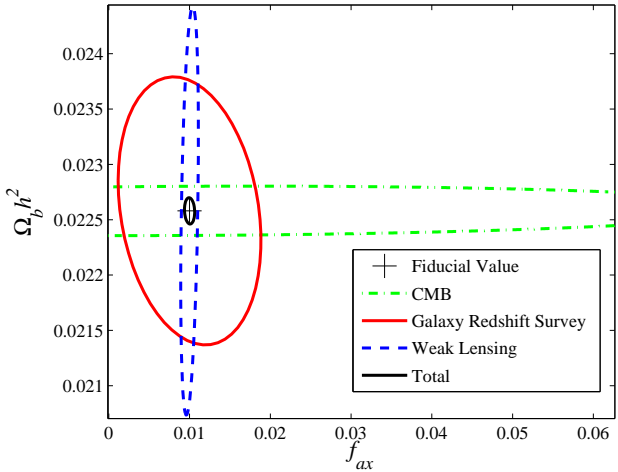}
\caption{Constraints on $\Omega_{b}h^{2}$ and $f_{ax}$, for $m_a=10^{-30}\unit{eV}$. We note a negative correlation from galaxy redshift surveys, whereas at this level of constraint, no correlation is visible in the CMB..}
\label{Ellipse_Ob_fAx_allObs} 
\end{figure}

Fig.~\ref{Ellipse_fAx_ns_AllObs} shows the expected positive correlation between constraints on $f_{ax}$ and constraints on $n_s$ from GRS. The constraints on $\Omega_b h^2$ and $f_{ax}$ are shown in Fig. \ref{Ellipse_Ob_fAx_allObs} show a negative correlation in GRS, and small positive correlation in WL. Both figures show no visible correlation in the CMB on the scale of the plot. The strong constraining power of the CMB on $\Omega_b h^2$ and $n_s$ relative to $f_{ax}$ breaks the expected correlations at this level.

For comparison, we also considered a cosmology as before, but with massless neutrinos. The constraints on $f_{ax}$ from all observables, along with the mass dependence, was approximately the same as without neutrinos.

\begin{figure}
\centering
\includegraphics[width=9cm]{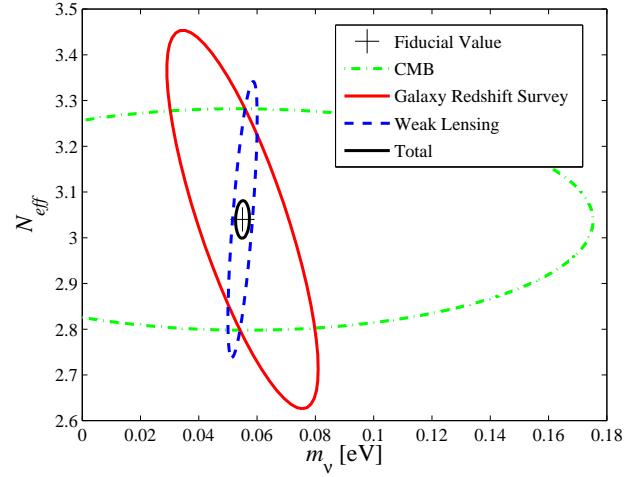}
\caption{Constraints on $N_{eff}$ and $m_{\nu}$ for the axion-free cosmology.     }
\label{Ellipse_Neff_mNu_fAx0} 
\end{figure}

We finally consider an axion-free cosmology, with massive neutrinos.  The Fisher ellipse for $N_{eff}$ and $m_{\nu}$ is shown in Fig. \ref{Ellipse_Neff_mNu_fAx0} for our different observables.

%
%
\section{Discussion and Conclusions}

Ultra-light axions and sub eV mass neutrinos are on the face of it very similar, when viewed in their effects on the matter power spectrum, and we seem hard pressed to distinguish them from $\Lambda$CDM, never mind one another, within the limits of current or future observations of galaxy clustering alone. There are in addition many other cosmological ingredients that can mimic various parts of such a signal and display degeneracies with axions and neutrinos. Despite this, we have seen that in fact quite tight bounds could be placed on the axiverse by combining many precision cosmological observations, in particular weak lensing tomography.

All the effects of an ultra-light species stem from two main sources: the effective removal of some dark matter from the background expansion while the field is frozen at redshifts greater than $z_{osc}$, and the suppression of structure due to the quantum pressure of the field below the scale $k_m$, an effect very similar to neutrino free streaming. This is because the axions are having purely gravitational effects. The effects of quantum pressure on the matter power spectrum, growth rate, and galaxy weak lensing can all be qualitatively understood in terms of the simple step picture provided by the fitting of $T_{ax}$. The effects on the CMB, and additional effects in the BAO can be understood simply from the transition in scalar field behaviour at $z_{osc}$.

An important positive correlation exists between the fraction in axions, $f_{ax}$, and the total content in CDM, $\Omega_c h^2$, since these ingredients have opposite effects in moving cosmological scales, such as equality and Silk damping. However, $f_{ax}$ can still be quite tightly constrained  and this degeneracy broken. In the case of GRS observables alone, a fraction in axions of 1\% can be distinguished from zero, independent of mass. Redshift information breaks the degeneracy with CDM by resolving growth rate effects that compliment those obtained from the power spectrum alone. Axions cause a distinctive scale dependent suppression of the growth rate at the scale $k_m$, in a manner which can be understood in order of magnitude as being due to the step in the power spectrum, $T_{ax}$. $\Omega_c h^2$ also controls the magnitude of the growth rate, so that measurements of growth and clustering together can constrain CDM and axions independently.

We have checked for dependence of our results when varying $k_{max}$ and the bias. A change in $k_{max}$ from $0.156$ to $0.146$ has very little effect on the GRS uncertainties on $f_{ax}$, as expected due to the axion effects occurring predominantly at much smaller $k$'s. However, introducing a redshift dependent bias of $b=0.6 (1+z)$ has a slightly larger effect, pushing the $1\sigma$ error to be \emph{just} larger than $f_{ax}=0.01$ from GRS alone. 

The \emph{mass independence} of axion constraints from galaxy redshift surveys and weak lensing can be important for the axiverse scenario, since a measurement will consequently constrain the existence of any and all species in the mass range we have investigated, where axions begin oscillating in the matter dominated era, $m_a\lesssim 10^{-28}\unit{eV}$. Such species can be resolved, at $1\sigma$, in quantities of around 1\% of the total DM content. The constraint is slightly worse when massive neutrinos are not present. The small dependence on neutrinos can be understood since neutrino mass adds more radiation at early times, and consequently moves the relevant scales, in this case making the axion effects slightly harder to resolve in the surveys considered. 

The correlation of axions with baryon content is more subtle, in that it is observable dependent. We expected a positive correlation in CMB observables: axions shift equality one way, baryons adding more matter, shift it the other, but the vastly different constraining power of the CMB between these observables did not make this visible. However, there is a negative correlation in GRS, where baryons have the competing effect of also suppressing structure. Similarly, we see a positive correlation between axions and the number of effective neutrino species. One normally thinks of extra relativistic energy density causing suppression of power in the CMB, but this is only true if all other parameters are altered so as to preserve equality and the angular horizon size. A simple increase in $N_{eff}$ actually boosts CMB power, and competes in an opposite manner to axions. If equality and horizon size are fixed in the presence of axions, the CMB is left unchanged: axions effect the CMB purely through the expansion rate, the change to which is only slight near $z_{osc}$, making the unlensed CMB a poor observable to look for axions.

Constraints on axions from the CMB are subtly dependent on axion mass. Although the axions effect the CMB purely gravitationally via the expansion rate, they do not simply ``switch on'' to DM behaviour out of nowhere, but contribute as a DE term prior to this. If the DE contribution is significant near to the surface of last scattering, then the CMB becomes more sensitive to these axions. This is true for the heaviest axions we have considered, since their masses cause slow roll to occur near to last scattering, and the axions transition through this while contributing of order $f_{ax}$ to the energy density. The constraints on this axion mass range coming from the CMB are thus related to well known CMB constraining power on EDE. This result is also consistent with the results of \cite{amendola2005}, where it is this mass range that is most strongly constrained by the CMB, with a sharp rise in uncertainty outside of this range when axions are either pure CDM or pure cosmological constant as far as the background expansion post recombination, and hence the CMB is concerned. However, axions are distinct from EDE in that they exit slow roll shortly after this, whilst EDE does not. An investigation into correlations between such axions and popular models of EDE would be interesting to pursue. 

In all cases, we have seen the tightest constraints coming from weak lensing, and just like GRS the strength of this constraint depends on the photometric redshift measurement i.e. on tomography. Lensing tomography allows another measurement of the growth rate, and the redshift evolution of the axion suppression of small scale convergence power can be resolved. We have checked that varying $\ell_{max}$ from $1900$ to $1500$ has very little effect on the uncertainty. This is, just as the case with GRS, expected since axions have their effects predominantly at much smaller $\ell$ (larger scales) than this.

Our major result has been to show that with current and next generation galaxy surveys alone it should be possible to unambiguously detect a fraction of dark matter in axions of the order of a few percent of the total.

In conclusion, we have seen that even such a simple ingredient as a light, non-interacting scalar field, such as the axion, can lead to interesting cosmology. The effects are sometimes subtle, and minor, making such an ingredient hard to spot. When these constraints are independent of the mass of the scalar field, they will be important to bound a large range of parameter space in the string axiverse. However, we have shown there is the potential to determine ultra-light axions as a distinct dark matter ingredient to high precision using future observations, in particular weak lensing tomography.  
\section*{Acknowledgements}
\vspace{-10pt}
DJEM thanks Joe Zuntz for being a constant help with getting to grips with CAMB in the earliest stages of this project, and Anthony Lewis for a helpful reply to a CAMB query. DJEM also thanks Ren\'{e}e Hlozek for many useful discussions. EM thanks Fergus Simpson, Will Percival and Yun Wang for help with the galaxy redshift survey Fisher matrices.  We thank Sudeep Das for the use of FisherCodes. We thank the STFC, BIPAC and Oxford Martin school for support.


\bibliographystyle{h-physrev3}
\bibliography{doddyoxford,EDbib,science}

\appendix

\section{Forecast Survey Details}
\label{appendix_survey_details}
In this appendix we include details of the surveys. The central redshift, volume, and $k_{\text{max}}$ for each of the 15 redshift bins in our galaxy redshift survey are given in Table~\ref{redshift_bins}.

\begin{table}[h]
\begin{center}
\begin{tabular}{|c|c|c|}
\hline
Central Redshift & Volume $(h^{-3}$Gpc$^3)$ & $k_{\text{max}} (h $Mpc$^{-1})$ \\
\hline
0.571 & 1.64 & 0.146\\
0.673 & 1.91 & 0.156\\
0.775 & 2.15 &0.166\\
0.877 & 2.36 & 0.189\\
0.979 & 2.53 & 0.20 \\
1.081 & 2.67 & `'\\
1.183  & 2.79 & $\downarrow$ \\
1.285  & 2.88 &\\
1.387  & 2.96 & \\
1.489 & 3.02&\\
1.591 & 3.07 & \\
1.693  & 3.12 &\\
1.795  & 3.13 &\\
1.897  & 3.15 &\\
1.999 & 3.16&\\
\hline
Total volume & 43.68& \\
Total galaxies & 43.68 million &\\
\hline
\end{tabular}
\end{center}
\caption{Details of the galaxy redshift survey model, designed to be similar to the Euclid galaxy redshift survey.  A constant galaxy bias of 1.7 and density of $1\times10^{-3} h^{3}$Mpc$^{-3}$ was assumed.}
\label{redshift_bins}
\end{table}

For the lensing survey, we define the redshift bins in Table~\ref{wl_redshift_bins}, and give the survey parameters in Table~\ref{wl_parameters}.

        \begin{table}[h]
        \begin{center}
                \begin{tabular}{|c|c|}
                \hline
                Lower limit & Upper limit \\
                \hline
                0 & 0.777 \\
                0.777 & 1.098 \\
                1.098 & 1.42 \\
                1.42 & 1.838 \\
                1.838& 3.0\\
	      \hline
                \end{tabular}
        \end{center}
        \caption{Details on the definition of the five redshift bins for weak
lensing, as defined by Eq.~\ref{eqn:niofz}, designed to contain the
same number of sources in each bin.}
        \label{wl_redshift_bins}
        \end{table}
        \begin{table}[h]
        \begin{center}
                \begin{tabular}{|c|c|}
                \hline
                        Parameter & Value \\
                        \hline
                        $f_{sky}$ & 0.5 \\
                        $\ell_{max}$ & 1900 \\
                        $\langle \gamma_{int}^2 \rangle^{1/2}$ & 0.22 \\
                        $\bar{n}$ & $4.18 \times 10^{8}\unit{sr}^{-1}$ \\
                        $z_0$ & 0.9 \\
                        $\sigma_z$ & $0.03(1+z)$\\
                     \hline  
                \end{tabular}
        \end{center}
        \caption{Summary of parameters for the weak lensing survey.}
        \label{wl_parameters}
        \end{table}

\end{document}